\documentclass{aa}
\usepackage{graphicx}
\usepackage{amssymb}
\usepackage{longtable}

\begin{document}

   \title{The brown dwarf population in the Chamaeleon~I cloud
                \thanks{Based on observations collected at the European
                Southern Observatory, La Silla, Chile}
        }

   \subtitle{}

   \author{B. L\'opez Mart\'{\i}\inst{1}\inst{2},
          J. Eisl\"offel\inst{2},
          A. Scholz\inst{2}
          \and R. Mundt\inst{3}
           }

   \offprints{B. L\'opez Mart\'{\i}}

   \institute{Observatori Astron\`omic de la Universitat de Val\`encia,
             Edifici d'Instituts d'Investigaci\'o, Pol\'{\i}gon La Coma,
             E-46980 Paterna, Spain\\
             \email{Belen.Lopez-Marti@uv.es}
             \and Th\"uringer Landessternwarte, Sternwarte 5, 
                D-07778 Tautenburg, Germany\\
                \email{jochen@tls-tautenburg.de, scholz@tls-tautenburg.de}
                \and 
                Max-Planck-Institut f\"ur Astronomie, K\"onigstuhl 17, 
                D-69117 Heidelberg, Germany\\
                \email{mundt@mpia-hd.mpg.de}}

   \date{Received; accepted}

   \abstract{We present the results of a multiband survey for brown dwarfs in
        the \object{Chamaeleon~I} dark cloud with the Wide Field Imager (WFI)
        camera at the ESO/MPG 2.2-m telescope on La Silla (Chile). The survey
        has revealed a substantial population of brown dwarfs in this southern
        star forming region. Candidates were selected from R, I and H$\alpha$
        imaging observations. We also observed in two medium-band filters, M855
        and M915, for the purpose of spectral type determination. The former
        filter covers a wavelength range containing spectral features
        characteristic of M-dwarfs, while the latter lies in a relatively
        featureless wavelength region for these late-type objects. A
        correlation was found between spectral type and (M855--M915) colour
        index for mid- to late M-type objects and early L-type dwarfs. With
        this method, we identify most of our object candidates as being of
        spectral type M5 or later. Our results show that there is no strong
        drop in the number of objects for the latest spectral types, hence
        brown dwarfs may be as abundant as low-mass stars in this region. Also,
        both kind of objects have a similar spatial distribution. We derive an
        index $\alpha = 0.6 \pm 0.1$ of the mass function in this region of
        dispersed star formation, in good agreement with the values obtained in
        other star forming regions and young clusters.
        Some of the brown dwarfs have strong H$\alpha$
        emission, suggesting mass accretion. For objects with published
        infrared photometry, we find that strong H$\alpha$ emission is related
        to a mid-infrared excess, indicative of the existence of a
        circumstellar disk. 

   }

\titlerunning{The brown dwarf population in the Chamaeleon~I cloud}
\authorrunning{L\'opez Mart\'{\i} et al.}
   \maketitle
%
\section{Introduction}

        Our understanding of substellar objects has considerably increased in
the last years. Recent surveys have revealed important populations of brown
dwarfs in young open clusters and star forming regions (e.g. Bouvier et al. 
\cite{bouvier98}; B\'ejar et al. \cite{bejar99}; Wilking et al. 
\cite{wilking99}; Lucas \& Roche \cite{lucas00}; Barrado y Navascu\'es et al. 
\cite{bn01}; Zapatero Osorio et al. \cite{zo02}; Lamm et al. in prep.). Still,
little is known about the properties of very young substellar objects 
(ages~$<10^8$~yr). 

        An open question is the way brown dwarfs form. Although it is commonly
accepted that they originate, like low-mass stars, from the gravitational
collapse of molecular cloud cores, their formation process is still poorly
understood. The number of observed brown dwarfs seems to vary  from one region
to another. For instance, some \object{Orion} clusters are rich in  substellar
objects (Lucas \& Roche \cite{lucas00}; B\'ejar et al. \cite{bejar99}), while
in \object{Taurus} only a few brown dwarfs are known so far (Brice\~no et al.
\cite{briceno98}; Mart\'{\i}n et al. \cite{martin01}). Hence, environmental
conditions may play an important role in the formation of brown dwarfs. 
Stellar winds from a nearby hot massive star could evaporate the envelopes of 
accreting cores before they reach a mass above the hydrogen burning limit. 
Another possibility is that brown dwarfs are stellar embryos whose growth 
stopped after they were dynamically ejected from small stellar systems 
(Reipurth \& Clarke \cite{reipurth01}; Bate et al. \cite{bate03}). Brown dwarfs
may also form like planets, in accretion disks around stars (e.g. Pickett et 
al. \cite{pickett00}). In this case, most of them would have been ejected from 
their parental systems to account for the so-called ``brown dwarf desert'' 
(e.g. Sterzik \& Durisen \cite{sterzik}). Another way to explain the lack of 
brown dwarf companions in wide orbits around stars is that they have migrated 
inwards due to their high mass (compared to planets). Eventually, they could 
remain in closer orbits, or continue their migration until they merge with the 
central star (Armitage \& Bonnell \cite{armitage02}).

        The Chamaeleon~I cloud is a favourable region for the study of the
formation of substellar objects, because of its proximity ($\sim$160~pc,
Whittet et al. \cite{whittet97}) and high galactic latitude 
($b\approx-16^{\circ}$), which decreases the effects of contamination by other
late-type stars in the Galaxy. It is the largest of the three dark clouds in 
the \object{Chamaeleon} complex (Gregorio Hetem et al. \cite{gh89}), and
apparently also the oldest. The cloud contains  more than 150 known young 
stars (e.g. Gauvin \& Strom \cite{gauvin92}; Feigelson et al. 
\cite{feigelson93}; Appenzeller \& Mundt \cite{appenzeller89}) with a mean age 
of about 3~Myr, most of them clustered in two cloud cores containing two 
intermediate mass stars, \object{HD~97048} and \object{HD~97300}. Chamaeleon~I 
has also been surveyed for brown dwarfs: Neuh\"auser \& Comer\'on 
(\cite{neuhauser98}) reported the discovery of the very first X-ray emitting 
brown dwarf, ChaH$\alpha$1, in Chamaeleon~I. In subsequent observations, 
another twelve bona-fide brown dwarfs and brown dwarf candidates (with masses 
between about 0.08 and 0.03$M_{\odot}$) were found in a small region around 
HD~97048 (Comer\'on, Rieke \&  Neuh\"auser 1999; Comer{\'o}n, Neuh{\"a}user \& 
Kaas 2000; hereafter \cite{comeron99} and \cite{comeron00}, respectively). 

        This paper presents a new large and deep survey for brown dwarfs in the
Chamaeleon~I cloud. In Sect.~\ref{sec:data} we describe the reduction and
photometry of our data. Our candidate selection criteria follow in
Sect.~\ref{sec:res}. A method for the spectral type identification of M and
early L-dwarfs from direct imaging is presented. We discuss our object
classification, and compare our identifications with available near- and
mid-infrared data. We then proceed to study the possible binarity of the
low-mass objects in Chamaeleon~I (Sect.~\ref{sec:bin}), their spatial 
distribution (Sect.~\ref{sec:dist}), their accretion indicators 
(Sect.~\ref{sec:accr}) and their mass function (Sect.~\ref{sec:chaimf}). The 
implications of our results for the proposed brown dwarf formation models are 
discussed in Sect.~\ref{sec:bdfor}. We summarize our conclusions in 
Sect.~\ref{sec:concl}.

%
\section{Observations and Data Processing}\label{sec:data}

\subsection{WFI Observations}\label{sec:wficha}

        The Wide Field Imager (WFI) is a focal reducer-type camera mounted at
the Cassegrain focus of the 2.2-m ESO/MPG telescope at La Silla (Baade et al.
\cite{baade98}). It consists of a mosaic of eight $2\times4$~K CCD detectors 
with narrow interchip gaps (the filling factor is 95.9\%). The detector has a 
field of view of about $34^{\prime}\times33^{\prime}$ and a pixel size of 
$0\farcs24$.

\begin{figure}
   \centering
  \includegraphics[width=9cm, angle=-90]{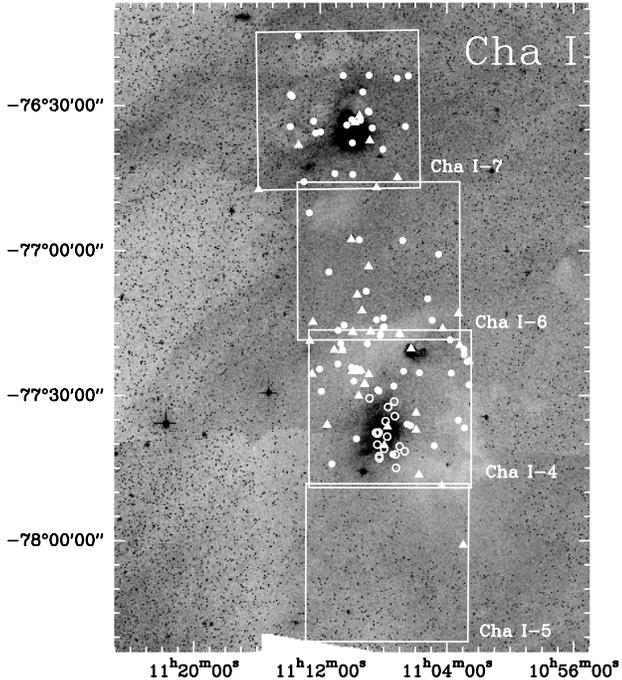} \hfill   
  \caption{DSS image of the Chamaeleon~I cloud showing the location of our 
                four WFI fields, ChaI-4 through 7. The open circles indicate 
		the positions of the  objects studied by \cite{comeron99} and
                \cite{comeron00}; the  filled circles and triangles, of our
                candidates with and without H$\alpha$ emission, respectively.}
              \label{fig:chaifield}
    \end{figure}

        In Chamaeleon~I, our survey consists of four WFI fields, covering an 
area of about 1.2~$\sq^{\circ}$ (see Fig.~\ref{fig:chaifield}). Two of these 
fields contain the intermediate mass stars HD~97048 and HD~97300, respectively,
around which most of the young low-mass stellar  population is clustered. The 
field previously surveyed for brown dwarfs (\cite{comeron99}; 
\cite{comeron00})
covered a region around HD~97048 that represents about 15\% of the total area 
of our ChaI-4 field. The other two fields lie in regions where no intermediate 
mass stars are present: ChaI-5 to the south of ChaI-4, and ChaI-6 in the region
between ChaI-4 and ChaI-7. We did not dithered between the fields, so some
objects might be missed because they fall in an interchip gap.

        The observations were carried out on three different nights in May/June
1999. We observed in the broad-band filters R and I, for the purpose of
candidate selection, and in a H$\alpha$ filter, to test the youth of our
objects. Because we wanted to attempt a photometric spectral classification
(see Sect.~\ref{sec:spt}), we also observed in two medium-band filters, 855/20
and 915/28\footnote{\footnotesize The transmission of these filters was
remeasured on December 16th 1999, and they were re-named 856/14 and 914/27,
respectively (see the WFI web site for details). However, we will keep the old
nomenclature throughout this paper.} 
(hereafter M855 and M915, respectively), centred inside and outside TiO and VO
absorption bands. These bands get deeper with later M-subspectral type, but
disappear in K- and L-objects.

        For each field and filter, we took three exposures with different
exposure times, in order to prevent the brightest T~Tauri stars (TTS) from
being saturated. For the H$\alpha$ filter only two different exposure times
were used. For the field ChaI-7 this observing programme could not be carried
out completely due to technical problems, and in this case we only have images
with short and intermediate exposure times in the R and I filters. The average
seeing varied from $\sim 1\arcsec$ for the field ChaI-7 to $\sim1\farcs7$ for
ChaI-5. The log of our observations is shown in Table~\ref{tab:log}.

        Bias exposures and skyflats were taken in each filter. For the 
broad-band photometry calibration, several Landolt fields (Landolt 
\cite{landolt92}) were observed in all nights. We also observed a list of
stars with known late spectral type (between K7 and L3) to use them as
standards for the spectral type calibration (see Sect.~\ref{sec:spt}). 
 
%
%
   \begin{table}
\footnotesize
   \centering
      \caption[]{Log of WFI observations in Chamaeleon~I}
         \label{tab:log}
         \begin{tabular}{p{0.2\linewidth}l c c c}
            \hline
            \hline
Date & Field & Filter & Exp. Times (s) \\
            \hline
28~May~1999 & ChaI-4 & Rc/162           & 600/60/5\\
28~May~1999 & ChaI-4 & Ic/lwp           & 600/30/5\\
28~May~1999 & ChaI-4 & H$\alpha$/7      & 600/15\\
28~May~1999 & ChaI-4 & 855/20           & 600/300/16\\
28~May~1999 & ChaI-4 & 915/28           & 600/100/8\\
\hline
28~May~1999 & ChaI-6 & Rc/162           & 600/60/5\\
28~May~1999 & ChaI-6 & Ic/lwp           & 600/30/5\\
28~May~1999 & ChaI-6 & H$\alpha$/7      & 600/15\\
28~May~1999 & ChaI-6 & 855/20           & 600/300/16\\
28~May~1999 & ChaI-6 & 915/28           & 600/100/ 8\\
\hline               
30~May~1999 & ChaI-7 & Rc/162           & 60/5\\
30~May~1999 & ChaI-7 & Ic/lwp           & 30/5\\
30~May~1999 & ChaI-7 & H$\alpha$/7      & 600/15\\
30~May~1999 & ChaI-7 & 855/20           & 600/300/16\\
30~May~1999 & ChaI-7 & 915/28           & 600/100/8\\
\hline
 3~Jun~1999 & ChaI-5 & Rc/162           & 600/60/5\\
 3~Jun~1999 & ChaI-5 & Ic/lwp           & 600/30/5\\
 3~Jun~1999 & ChaI-5 & H$\alpha$/7      & 600/15\\
 3~Jun~1999 & ChaI-5 & 855/20           & 600/300/16\\
 3~Jun~1999 & ChaI-5 & 915/28           & 600/100/8\\
\hline
\end{tabular}
\end{table}
%

\subsection{Data Reduction and Photometry} \label{sec:red}

        The image processing (data reduction and  photometry) was run with the
MSCRED and CCDRED packages in the IRAF environment.\footnote{\footnotesize 
IRAF is distributed by the National Optical Astronomy  Observatory (NOAO), 
which isoperated by the Association of Universities for  Research in Astronomy,
Inc. under contract to the National Science  Foundation.}  
Due to the large field of view, a standard reduction of WFI images (bias 
subtraction and flat field division) is not enough to obtain good photometry
results: First, because the illumination of the sky flat fields and the 
science images is not equivalent, there remains an illumination gradient in the
background of our object exposures after the flatfielding. To correct for  
this effect, the image has to be divided through this gradient or 
\emph{illumination mask}. Second, due to the long exposure times, a fringing
pattern appears. This effect is especially strong in  the I band. To correct
for it, one has to subtract an individually  scaled \emph{fringe mask} from
each science image. The fringes cannot be completely removed, however, because
their intensity varies from one chip to  the other and even between different
areas of the same chip. On the other hand, in the R and H$\alpha$ images, as
well as in most of the images with short exposure times, the fringing effect 
is negligible, and therefore we only corrected for the illumination gradient in
those cases. Both the illumination and the fringe masks were created by
combination of the science exposures. 

For the astrometry, we followed the standard method proposed in Francisco 
Vald\'es' \emph{Guide to the NOAO Mosaic Data Handling 
Software}.\footnote{This guide is available from: \\
{\tt http://iraf.noao.edu/scripts/irafhelp?mscguide}}
Stars from the Guide Star Catalogue were selected as reference for the 
astrometric calibration (Morrison et al. \cite{morrison01} and references 
therein). The astrometric error is in general not larger than 1$\arcsec$.

        Object search was then performed on each individual chip of the long
exposures using SExtractor (Bertin \& Arnouts \cite{bertin96}). After several
tests, we concluded that this software worked better than DAOFIND, leading to
fewer spurious detections. 

        For the photometry, the DAOPHOT  package was used (Stetson
\cite{stetson87}). Since the DAOPHOT tasks only work on individual CCDs, a
special IRAF procedure was written for the WFI images. We first performed
aperture photometry with the PHOT task, and then PSF-photometry using the PSF 
and ALLSTAR tasks, regardless of the crowdedness of the field. In this way we 
intended to measure the faintest components of multiple systems. This did not 
prevent some very faint objects to be merged with their much brighter 
neighbours, however, although we allowed only for a very small merging radius. 
This problem affects to objects up to separations of about 4$\arcsec$ 
depending on the relative brightness of the objects. Given that the fraction 
of faint objects lost due to this problem is small (less than 3\%) in such an 
uncrowded region, we decided not to further consider these objects in our 
study. Only in a few interesting cases (see Sect.~\ref{sec:bin}), aperture 
photometry was performed on the subtracted images.

On most of the short exposures only aperture photometry was performed, because 
of the difficulty of fitting a suitable PSF. This was also the case of the 
standard stars (exposure times of 5--10\,s).

\subsection{Photometric Calibration} \label{sec:cal}

        The two broad-band filters were calibrated with Landolt
(\cite{landolt92}) standard stars. Absolute magnitudes were computed using the
relations:

\begin{center}
\begin{equation}
\label{eq:rcal}
R = r + B_1 + B_2 \cdot X_R + B_3 \cdot (r-i)
\end{equation}
\begin{equation}
\label{eq:ical}
I = i + A_1 + A_2 \cdot X_I + A_3 \cdot (r-i)
\end{equation}
\end{center}

\noindent 
where $r$ and $i$ are the instrumental magnitudes, $X_R$ and $X_I$ denote the
measured airmasses, and $A_k$ and $B_k$ are the coefficients obtained from a
linear fit using the standard instrumental and absolute magnitudes.

        In the first two nights (May 28th and May 30th), five Landolt fields
containing more than 60 standard stars were observed. Their airmasses range 
from 1.15 to 2.90. The fits for these nights have a standard deviation of 
about 5\%. In the fourth night (June 3rd), only two Landolt fields were taken, 
although one of them was observed twice at different airmass. The number of 
stars used for the fit (the stars in the repeated field were used twice) was 
29, with airmasses between 1.14 and 1.95. The standard deviation is about 6\%.

        The resulting coefficients are somewhat different for each night (see
Table~\ref{tab:coeff}). In particular the colour term for the I band, $A_3$, 
is relatively high. This is not surprising, because the shape of the 
transmission curve for this filter differs notably from a standard Cousins I 
filter, especially at shorter wavelengths. On the other hand, the R filter 
shows a good match with a Cousins R filter, and in this case the colour term is
so small that it can be set to zero (see the \emph{WFI User Manual} for 
details).

   \begin{table}
     \footnotesize
     \centering
      \caption[]{Calibration coefficients for the WFI I and R filters.}
         \label{tab:coeff}
     $$ 
         \begin{tabular}{p{0.2\linewidth} l c c c c c c}
            \hline
            \hline
Night & $A_1$ & $A_2$ & $A_3$ & $B_1$ & $B_2$ & $B_3$\\
            \hline
May 28th & -1.38 & -0.13 & 0.23 & -0.21  & -0.24 & 0.00 \\
May 30th & -1.47 & -0.03 & 0.25 & -0.58  &  0.00 & 0.00 \\
June 3rd & -1.19 & -0.19 & 0.32 & -0.18  & -0.23 & 0.00 \\
            \hline
         \end{tabular}
     $$ 
   \end{table}
%

   \begin{table*}[t]
  \centering
 \scriptsize
      \caption[]{\footnotesize Photometry and spectral types of the well known objects in 
                Chamaeleon~I\,$^{\mathrm{ab}}$}
         \label{tab:nc}
         \begin{tabular}{p{0.11\linewidth}c c c c c c c l l l l}
            \hline
            \hline
\small
Name & $\alpha$ (2000) & $\delta$ (2000) & R & I & H$\alpha$ & M855 & M915 
& SpT$^{\mathrm{c}}$ & SpT$^{\mathrm{d}}$ & Classification\\
            \hline
 \object{CHXR~78C}        & 11 08 54.7 &  -77 32 12.3 & 17.50 & 15.12 &  17.23 & 18.92 &  18.22  & M5.5  & M5.5  & star \\
 \object{CHXR~21}         & 11 07 11.8 &  -77 46 37.8 & 15.28 & 13.46 &  15.04 & 17.24 &  16.86  & K7    & $<$M4 & star \\
 \object{CHXR~22}         & 11 07 13.7 &  -77 43 50.3 & 17.21 & 14.96 &  16.94 & 18.65 &  18.14  & K7+M2 & M4    & star \\
 \object{CHXR~74}         & 11 06 57.8 &  -77 42 11.2 & 15.92 & 13.91 &  15.59 & 17.70 &  17.20  & M4.5  & M4    & star \\
 \object{HM~19}           & 11 08 16.9 &  -77 44 37.8 & 15.12 & 13.31 &  14.88 & 17.18 &  16.76  & M3.5  & $<$M4 & star \\
 \object{Sz~23}           & 11 07 58.5 &  -77 42 42.0 & 16.55 & 14.64 &  15.59 & 18.43 &  18.03  & M2.5  & $<$M4 & star \\
 \object{ChaH$\alpha$~1}  & 11 07 17.2 &  -77 35 53.5 & 19.45 & 16.62 &  18.95 & 20.54 &  19.40  & M7.5  & M9    & BD \\
 \object{ChaH$\alpha$~2}  & 11 07 42.9 &  -77 34 00.2 & 17.96 & 15.42 &  17.15 & 19.19 &  18.44  & M6.5  & M6    & trans. obj. \\
 \object{ChaH$\alpha$~3}  & 11 07 52.7 &  -77 36 57.6 & 17.80 & 15.26 &  17.61 & 19.15 &  18.30  & M7    & M6.5  & trans. obj. \\
 \object{ChaH$\alpha$~4}  & 11 08 19.4 &  -77 39 17.8 & 17.08 & 14.74 &  16.85 & 18.63 &  17.89  & M6    & M6    & trans. obj. \\
 \object{ChaH$\alpha$~5}  & 11 08 24.6 &  -77 41 48.1 & 17.56 & 15.02 &  17.36 & 18.89 &  18.06  & M6    & M6.5  & trans. obj. \\
 \object{ChaH$\alpha$~8}  & 11 07 46.5 &  -77 40 09.6 & 18.37 & 15.78 &  18.22 & 19.67 &  18.79  & M6.5  & M7    & BD \\
 \object{ChaH$\alpha$~9}  & 11 07 19.2 &  -77 32 52.4 & 20.6; & 17.65 &  19.95 & 21.30 &  20.46  & M6    & M6.5  & trans. obj. \\
 \object{ChaH$\alpha$~10} & 11 08 24.5 &  -77 39 30.7 & 19.86 & 17.25 &  19.74 & 21.09 &  20.22  & M7.5  & M7    & BD \\
 \object{ChaH$\alpha$~11} & 11 08 29.7 &  -77 39 20.4 & 20.4; & 17.72 &  19.89 & 21.59 &  20.65  & M8    & M7    & BD \\
 \object{ChaH$\alpha$~12} & 11 06 38.4 &  -77 43 09.7 & 18.78 & 16.09 &  18.45 & 19.96 &  18.96  & M7    & M8    & BD \\
 \object{ChaH$\alpha$~13} & 11 08 17.5 &  -77 44 12.5 & 16.95 & 14.52 &  16.64 & 18.41 &  17.59  & M5    & M6.5  & trans. obj. \\
                     
\hline
\end{tabular}
\begin{list}{}{}
\item[$^{\mathrm{a}}$] H$\alpha$, M855 and M915 magnitudes are instrumental
                        magnitudes (see text).
\item[$^{\mathrm{b}}$] Photometric errors: blank: 0.05; semicolon: 
                        0.1~mag; colon: 0.2~mag 
\item[$^{\mathrm{c}}$] Spectral types from \cite{comeron00} (brown dwarfs and
                        brown dwarf candidates) and \cite{comeron99} (stars).
\item[$^{\mathrm{d}}$] Spectral types from this work. The estimated errors are
                        1 subclass in the range M6-M9 and 2 subclasses in the
                        range M4-M6 (see text).
\end{list}
   \end{table*}

        The main error source in this calibration is the fact that it comes
from a \emph{global} fit for all the CCD chips. Recently, Alcal\'a et al.
(\cite{alcala02}) showed that, for the WFI, chip-to-chip photometric variations
can be as high as 3\% in the broad-band filters and 5\% in medium-band filters,
because of different zero-points for the individual CCDs. With no doubt, a
better result could be obtained by fitting the stars in each frame separately.
This was not possible in our case, however, because some CCD chips contained
few or no standard stars (especially those in the corners). Future 
observations should take several exposures dithering the fields, so that enough
stars are observed on each CCD.

We also note that the Landolt stars used for the photometric calibration span 
a range in (R--I) colours of about --0.1 to 1.5, while our objects of interest 
have colours in the range 0.5 to 4 (see Sect.~\ref{sec:bdchai}). Hence, 
our colours may not be exactly in the Cousins system, especially for the 
reddest objects with the largest colour terms. This is a general problem 
that affects all studies of very red objects, as long as no set of very 
red standard stars is available. For the WFI colours, comparison with the 
predictions of the Lyon models (Baraffe et al. \cite{baraffe98}; Chabrier et 
al. \cite{chabrier00}) shows deviations of up to 0.5~mag for very red objects 
objects (R--I=2.5), while there is good agreement for R--I=1.5 (see also 
Eisl\"offel et al., in prep.).

        Apart from this, our photometry shows a systematic error of unknown
origin: Using the objects present in the images with different exposure times
to estimate the error in our photometry, we found a systematic offset between
them. In general, the resulting luminosities are brighter by a constant amount
(typically between 0. and 0.06~mag, though as high as $\sim 0.1$~mag in the
worst case) for longer exposure  times. This effect is reproduced with various
photometry packages and with different parameters. It is  present already in
the raw data (i.e., previous to the reduction), and does not show any obvious
correlation with the observed field, the filter used or the measured
extinction. Observing conditions are not likely to have caused it, because the
images were taken one after another, and the effect is always in the same
direction for all images in all photometric nights. Extensive tests by the ESO
staff have verified that the effect is not caused by a WFI shutter malfunction
leading to systematically short or variable integration times. It seems also
improbable that it is caused by detector saturation, because the objects 
appear brighter, not fainter, with longer exposure times. 

        Therefore, to homogenize our photometry, we have determined the offsets
between the three different integration times, and have shifted the short and
long integration times to the system of the medium integration. That way, we
should have minimized this systematic error, since the medium integration times
for our objects are not very different from those of the standard stars. We did
not choose the shortest exposures as reference because they contained only a
few objects which could be used to estimate the exposure time offsets.

        Since our observed fields in Chamaeleon~I are slightly overlapping, we
checked for double detections of objects. We found that the photometry results
in the overlapping region between two fields are consistent within the
estimated errors. Comparison with DENIS I photometry for  previously known
objects  in our field ChaI-4 (Neuh\"auser \& Comer\'on \cite{neuhauser99})
gives differences of about 0.2~mag. Hence, our results are marginally
consistent with the DENIS measurements, whose errors for faint objects can get
as high as 0.2~mag for I$\simeq$18~mag (Epchtein et al. \cite{epchtein97}). A
larger difference is found when comparing our RI photometry with the results of
\cite{comeron99} and  \cite{comeron00}: Our results yield systematically
fainter magnitudes by about 0.4 mag in average in both passbands. This is not
surprising, given that the photometry of \cite{comeron99} and \cite{comeron00}
was calibrated with few secondary standards in the observed field, which are
probably variable, and without a colour correction. Because our error seems to
be approximately equal for both the R and I photometry, our R$-$I colours
should be very little affected by the exposure time effect.

        Because no standard stars are available for the medium-band filters, no
absolute calibration was possible for the H$\alpha$, M855 and M915 photometry, 
and only instrumental  magnitudes will be considered for the data in these 
filters. We applied a standard correction for atmospheric extinction, using the
extinction coefficients of the R and I band for the H$\alpha$ and the two M
filters, respectively.  For a better understanding of the emission properties, 
the H$\alpha$ magnitudes were then shifted so that the main locus of the 
objects in each surveyed field corresponds to H$\alpha-$R=0 in a (H$\alpha$, 
H$\alpha$--R) colour-magnitude diagram (see Sect.~\ref{sec:res} below). 

To estimate the completeness limits of our survey, we constructed the number 
distribution of the found objects as a function of their measured magnitudes 
for our four WFI fields. These distributions peak for values of R$\simeq$20~mag
and I$\simeq$19~mag, except for the field ChaI-4, where these values are 
shifted $\lesssim$1~mag towards fainter objects for both filters. There is a 
sharp drop in the number of objects for R$\simeq$23~mag (22~mag for ChaI-7, due
to the shorter exposure times) and I$\simeq$21~mag, respectively. Inspection of
the corresponding images shows that objects fainter than these values are most 
likely noise fluctuations. Hence, we consider general detection limits of about
23~mag and 21~mag for the R and I filters, respectively, and completeness 
limits of R$\simeq$20~mag and I$\simeq$19~mag, although the survey might be 
complete for fainter objects in  the area covered by the field ChaI-4.

%
\section{Survey Results}\label{sec:res}

\subsection{Brown Dwarf Candidates} \label{sec:bdchai}

%
   \begin{figure}
   \centering
   \includegraphics[width=6cm, angle=-90, bb= 50 70 550 775]{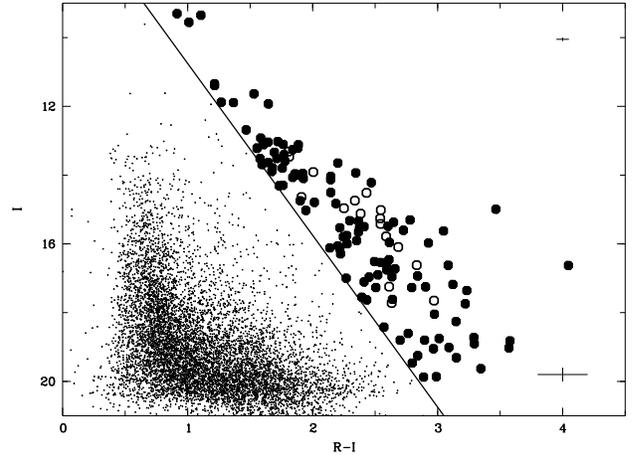} \hfill
     \caption{(I, R--I) colour-magnitude diagram for our surveyed field in
                Chamaeleon~I. The previously identified low-mass stars and
                brown  dwarfs from \cite{comeron99} and \cite{comeron00} (open
                symbols) define an empirical isochrone towards redder colours,
                where we select our new candidates (solid symbols). The crossed
                lines indicate the average errors.}
         \label{fig:richai}
    \end{figure}

        At an early stage of their evolution, brown dwarf candidates in star
forming regions are easily  identified in (I, R--I) colour-magnitude diagrams.
Fig.~\ref{fig:richai} shows our diagram for Chamaeleon~I. In the already
surveyed region around the star HD~97048, we recover all the previously known
objects from \cite{comeron99} and \cite{comeron00} (except for two of them 
that happened to fall in the interchip gaps): They lie in the region 
I$\lesssim$14 and are clearly separated from background objects towards redder
(R--I) colours. Our optical photometry for these objects is summarized in
Table~\ref{tab:nc}. These previously known objects have an estimated age of
about 1-3~Myr, and define an empirical isochrone in the colour-magnitude
diagram, which can be used to select candidates in a much larger field. Such an
isochrone contains information about the distance and reddening of the objects
in the dark cloud. Moreover, by using an empirical rather than a theoretical
isochrone, we avoid the uncertainties still present in the evolutionary models
for very young ages (see Baraffe et al. \cite{baraffe02} for a critical
discussion).  

        Using such an isochrone, we find 113 candidates members of Chamaeleon~I
in the same region of the diagram and its prolongation towards fainter sources.
The solid line in Fig.~\ref{fig:richai} is a fit on the brown dwarfs and brown 
dwarf candidates from \cite{comeron00}, shifted to allow for the spread 
observed in the previously known objects and for the estimated photometric 
errors. Our selected candidates cover the magnitude range down to 
I $\simeq$ 20~mag. The objects tend to be clustered around the cloud cores (see
Fig.~\ref{fig:chaifield}). The region most devoid of candidate objects lies 
towards the south of the surveyed area, where only one object is found in the 
zone of interest of the diagram. 

        Given the proximity of Chamaeleon~I ($\sim$160~pc), we do not expect a
significant contamination from foreground objects among our candidates.
Moreover, due to its high galactic latitude ($b \sim -16$) and to extinction
through the dark cloud, we should not be able either to see many background
objects. To estimate the amount of contaminating objects in our field, we made
use of the simulations of stellar population synthesis in the Galaxy by Robin
\& Crez\'e (\cite{robin86})\footnote{\footnotesize The simulations can be
generated on-line at this URL:
\tt{http://www.obs-besancon.fr/www/modele/modele\_ang.html}}. 
Within a solid angle of 1 square degree in the direction of the Chamaeleon
Complex, these simulations yield a number of 5 to 7 objects in the region of
the (I, R-I) colour-magnitude diagram around a 1~Myr isochrone from the
evolutionary models of Chabrier et al. (\cite{chabrier00}). Since we find more
than hundred objects in the same region of the diagram, we are confident that
about 95\% of our candidates are true members of Chamaeleon~I.

We note that our survey also includes a number of detections in the I-band 
that have no counterpart in R. Most of them lie near the lower magnitude 
limit of our survey with values of I $=$ 20-21. These objects may be very 
red, but because of their non-detection in R 
nothing can be said about their nature.

   \begin{figure}
   \centering
  \includegraphics[width=6cm, angle=-90, bb= 50 70 550 775]{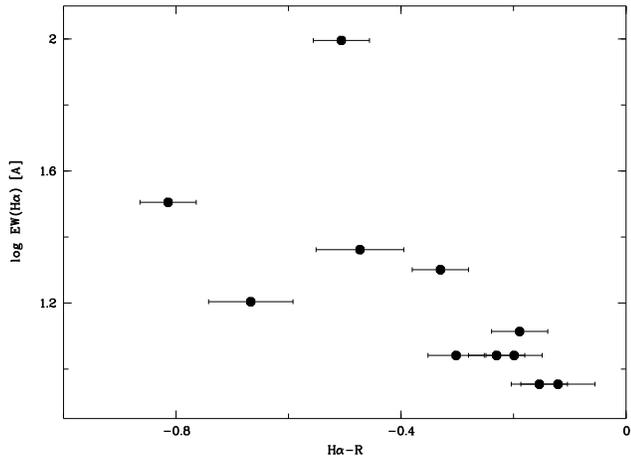} 
      \caption{Logarithm of the H$\alpha$ equivalent width versus the 
      (H$\alpha$--R) colour index for the objects in Chamaeleon~I with 
      measured equivalent width. The data are taken from \cite{comeron99} 
      for the low-mass stars, and from \cite{comeron00} for the brown dwarfs 
      and brown dwarf candidates.}
         \label{fig:eqwi}
   \end{figure}

        According to the theoretical models (Baraffe et al. \cite{baraffe98}; 
Chabrier et al. \cite{chabrier00}), if our candidates are indeed members of 
the star forming  region, they should reach down to very low masses, extending 
down to the deuterium burning limit for the faintest objects in our sample. One
way to confirm cloud membership is the detection of H$\alpha$  emission, which 
is usually considered as an indicator of youth. All the objects from 
\cite{comeron99} and \cite{comeron00} indeed show the H$\alpha$ line in 
emission in their spectra.

The amount of emission is usually quantified by the measurement of the 
H$\alpha$ equivalent width EW(H$\alpha$). Fig.~\ref{fig:eqwi} shows a plot of 
$\log \mathrm{EW(H\alpha)}$ versus our measured (H$\alpha$--R) colour index 
for the previously known brown dwarf and brown dwarf candidates. The values of 
EW(H$\alpha$) are taken from \cite{comeron00}.  With the notable exception of 
ChaH$\alpha$1, that will be later discussed in Sect.~\ref{sec:hal}, a tendency 
of smaller equivalent width towards higher colour index can be clearly seen. 
Of course, one has to be careful with such estimations, because the equivalent 
width measurements of \cite{comeron00} were performed at an epoch different 
from our optical observations, and these objects are known to have variable 
H$\alpha$ emission. However, still the (H$\alpha$--R) colour can be used as a 
rough estimate of the H$\alpha$ emission.

%
   \begin{figure}
   \centering
   \includegraphics[width=6cm, angle=-90, bb= 50 70 550 775]{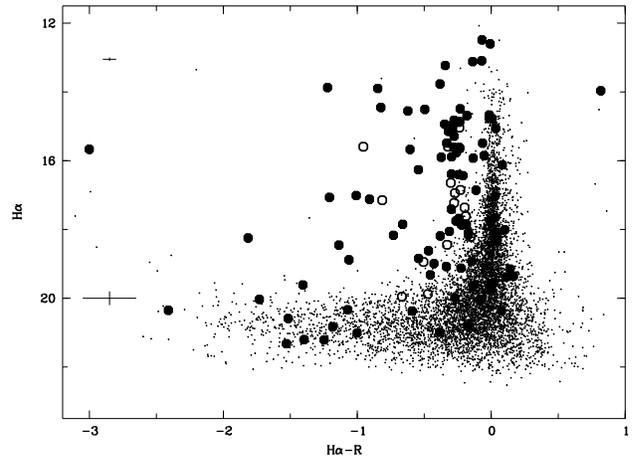} 
\hfill
       \caption{(H$\alpha$, H$\alpha$--R) colour-magnitude diagram for our
                surveyed field in Chamaeleon~I. Symbols as in 
                Fig.~\ref{fig:richai}. Objects with H$\alpha$ emission are 
                placed in the left part of the diagram.}
        \label{fig:harchai}
    \end{figure}

In Fig.~\ref{fig:harchai}, we present a (H$\alpha$, H$\alpha$--R) 
colour-magnitude diagram of the objects in our field. The objects from 
\cite{comeron99} and \cite{comeron00} are found in the left part of the
diagram, having H$\alpha$--R colours less than --0.1. Most of our bright
candidates are also located in this region of the diagram, and thus have
H$\alpha$ in emission.  Unfortunately, this cannot be clearly stated for most
of the faintest  objects, due to the large photometric errors in the lower part
of our diagram. However, some of them have such extreme measured
(H$\alpha$--R)  colours that they are likely H$\alpha$ emitters. Moreover, many
of them also show an infrared excess (see Sect.\ref{sec:ir}), another indicator
that they are  probably young. Thus, all the candidates with 
H$\alpha$--R$<$-0.1 in Fig.~\ref{fig:harchai} are considered probable
Chamaeleon~I members. In this way, we retain 69 objects. Of course, some of the
candidates without apparent H$\alpha$ emission will also belong to the cloud,
but this cannot be proved without spectra. 

        We later added three other objects to this list: ChaI~608, in spite of
being slightly too blue for our initial selection, is the probable optical
counterpart of the X-ray source CHXR~30 (Feigelson et al. \cite{feigelson93}).
ChaI~404 has no clear H$\alpha$ emission, but it could be the optical
counterpart of an ISOCAM source (see Sect.~\ref{sec:ir}). ChaI~616 has a 
reported infrared excess (see Sect.~\ref{sec:ir}). The coordinates and 
cross-identifications of all 72 objects are listed in Table~\ref{tab:chaiphot}.

%
   \begin{table*}[ht]
  \centering
  \scriptsize
      \caption[]{\footnotesize Candidate low-mass members of Chamaeleon~I$^{\mathrm{ab}}$}
         \label{tab:chaiphot}
         \begin{tabular}{p{0.07\linewidth}c c c c c c c l l l l}
            \hline
            \hline
Name & $\alpha$ (2000) & $\delta$ (2000) & R & I & H$\alpha$ & M855 & M915 & SpT$^{\mathrm{c}}$ & Classification & Other Names$^{\mathrm{d}}$ \\
            \hline
 \object{ChaI~401} & 11 11 59.3 & -77 30 38.2 &  22.9: &  19.9; &  21.3: &  23.2: &  22.9; &  L1.5 & BD cand.&  \\
 \object{ChaI~403} & 11 10 54.1 & -77 25 01.2 &  21.0; &  18.05 &  19.6; &  21.66 &  20.83 &  M6.5 & trans. obj. &  \object{DENIS-P J1110.9-7725}\\
 \object{ChaI~404}$^{\mathrm{e}}$ & 11 10 37.0 & -77 22 13.8  &  19.70 &  16.62 &  19.7; &  20.23 &  19.36 & M6.5 & BD cand. & \object{ISO 250}\,$^{\mathrm{f}}$ \\
 \object{ChaI~405} & 11 10 35.4 & -77 22 05.9 &  18.33 &  15.60 &  18.16 &  19.22 &  18.53 &  M5.5 & star & \object{ISO 250}\,$^{\mathrm{f}}$  \\
 \object{ChaI~406} & 11 10 42.0 & -77 20 48.8 &  20.06 &  17.27 &  18.25 &  21.06 &  20.06 &  M8 (M2) & star &  \object{ISO 252} \\
 \object{ChaI~408} & 11 10 51.4 & -77 18 04.0 &  16.65 &  14.50 &  16.41 &  18.19 &  17.70 &  M4 & star &  \\
 \object{ChaI~410} & 11 08 19.0 & -77 30 41.4 &  19.42 &  16.90 &  18.99 &  20.74 &  19.85 &  M7 (M5.5) & BD cand. &  \object{ISO 138} \\
 \object{ChaI~411} & 11 08 22.9 & -77 30 28.4 &  18.02 &  15.65 &  17.01 &  19.42 &  18.75 &  M5 & star &  \object{KG 45}, \object{ISO 143} \\
 \object{ChaI~412} & 11 09 53.9 & -77 28 37.2 &  21.4: &  18.26 &  20.33 &  21.98 &  20.87 &  M9 (M4.5) & star  &  \object{GK 30}, \object{ISO 220} \\
 \object{ChaI~417} & 11 09 48.0 & -77 26 29.8 &  18.27 &  16.00 &  17.07 &  19.61 &  19.12 &  M4 & star & \object{Baud 43}, \object{KG 101}, \\
 & & & & & & & & & & \object{ISO 207} \\
 \object{ChaI~418} & 11 09 43.1 & -77 25 58.4 &  18.90 &  15.97 &  18.17 &  19.61 &  18.78 &  M6.5 & trans. obj.  & \object{C7-1}, \object{KG 95}, \object{ISO 200}  \\
 \object{ChaI~419} & 11 10 02.5 & -77 25 45.9 &  21.0; &  17.74 &  20.4: &  21.32 &  20.44 &  M7 & BD & \\
 \object{ChaI~420} & 11 08 13.0 & -77 19 13.4 &  22.1: &  19.03 &  20.7: &  22.5; &  21.72 &  M9-9.5 (M1) & star  & \object{KG 37}, \object{ISO 131}, \\
& & & & & & & & & & \object{DENIS-P J1108.2-7719} \\
 \object{ChaI~422} & 11 03 42.5 & -77 26 52.8 &  18.57 &  15.96 &  18.19 &  19.75 &  18.87 &  M7 & BD &  \object{ISO 28} \\
 \object{ChaI~423} & 11 02 42.5 & -77 24 25.2 &  17.71 &  15.34 &  17.41 &  19.20 &  18.47 &  M6 & trans. obj. & \\
 \object{ChaI~424} & 11 02 54.5 & -77 22 57.0 &  22.5: &  19.3; &  21.2: &  23.3: &  22.1; &  M8.5 & BD & \\
 \object{ChaI~425} & 11 03 48.3 & -77 19 57.4 &  16.67 &  14.22 &  16.39 &  18.12 &  17.20 &  M7.5 & BD & \object{Hn2}, \object{ISO 32}, \object{DENIS 19} \\
 \object{ChaI~427} & 11 03 11.7 & -77 36 36.7 &  22.0: &  18.72 &  21.0: &  22.4; &  21.48 &  M9 & BD &  \object{DENIS-P J1103.2-7736} \\
 \object{ChaI~428} & 11 02 47.4 & -77 38 08.5 &  22.6: &  19.0; &  21.2: &  22.4; &  21.74 &  L0 (M1) &  & \object{DENIS-P J1102.8-7738} \\
 \object{ChaI~429} & 11 07 24.8 & -77 43 49.4 &  19.77 &  16.93 &  19.64 &  20.67 &  19.77 &  M7 & BD & \object{KG 10} \\
 \object{ChaI~432} & 11 06 15.9 & -77 37 50.5 &  19.07 &  16.46 &  18.92 &  20.02 &  19.50 &  M4 & star  & \object{ISO 76}, \\
& & & & & & & & & & \object{DENIS-P J1106.3-7737} \\
 \object{ChaI~433} & 11 06 29.3 & -77 37 33.6 &  18.26 &  15.91 &  18.09 &  19.56 &  19.05 &  M4 & star  & \object{CHXR73}, \object{ISO 78}  \\
 \object{ChaI~434} & 11 09 45.6 & -77 40 34.0 &  18.02 &  15.37 &  17.75 &  19.22 &  18.37 &  M7 & BD & \object{KG 96}, \object{ISO 201} \\
 \object{ChaI~436} & 11 11 22.8 & -77 45 43.6 &  21.8: &  18.75 &  20.0; &  22.6; &  21.50 &  M8.5 & BD &  \\
 \object{ChaI~438} & 11 12 04.0 & -77 26 01.5 &  18.51 &  16.29 &  17.85 &  20.08 &  19.46 &  M5 & star  & \object{ISO 282} \\
 \object{ChaI~441} & 11 09 01.7 & -77 20 53.1 &  21.0: &  18.42 &  20.8: &  22.1; &  21.63 &  L1 & BD & \object{ISO 171} \\
 \object{ChaI~448} & 11 07 21.0 & -77 29 41.0 &  15.10 &  13.22 &  14.86 &  17.11 &  16.59 &  M4 & star &  \object{ISO 99} \\
 \object{ChaI~449} & 11 05 43.6 & -77 26 52.3 &  15.85 &  13.65 &  15.61 &  17.54 &  16.84 &  M6 & trans. obj.  & \object{CHXR15}, \object{ISO 65} \\
 \object{ChaI~450} & 11 06 44.1 & -77 26 35.0 &  14.87 &  13.11 &  14.69 &  16.94 &  16.56 & $<$M4 & star & \object{T22}, \object{UX Cha}, \object{ISO 80}, \\
& & & & & & & & & & \object{DENIS 38} \\
 \object{ChaI~452} & 11 06 59.7 & -77 18 54.3 &  14.74 &  13.02 &  13.90 &  16.72 &  16.49 & $<$M4 (M1.5) & star  & \object{T23}, \object{UY Cha} \\
 \object{ChaI~453} & 11 02 33.3 & -77 29 13.6 &  14.72 &  13.11 &  14.49 &  17.03 &  16.69 & $<$M4 (M3) & star  & \object{CHXR71}, \object{Hn1}, \\
& & & & & & & & & & \object{ISO 4}, \object{DENIS 16}  \\
 \object{ChaI~456} & 11 02 55.7 & -77 21 51.5 &  15.17 &  13.40 &  14.56 &  17.35 &  16.77 &  M5 (M0.5) & star  & \object{T12}, \object{Sz10}, \object{DENIS 17}  \\
 \object{ChaI~457} & 11 04 43.1 & -77 41 57.7 &  15.82 &  13.96 &  15.49 &  17.82 &  17.36 &  M4 & star  & \object{B18}, \object{DENIS 26}  \\
 \object{ChaI~601} & 11 12 31.6 & -76 53 35.2 &  19.35 &  16.76 &  19.12 &  20.75 &  19.72 &  M8 & BD &  \\
 \object{ChaI~602} & 11 09 29.7 & -76 59 18.6 &  18.04 &  15.79 &  17.85 &  19.67 &  19.01 &  M5 & star &  \\
 \object{ChaI~603} & 11 04 57.6 & -77 15 57.2 &  16.65 &  14.74 &  16.43 &  18.51 &  18.05 &  M4 & star  & \object{T16}, \object{Sz13}, \object{ISO 55}, \\
& & & & & & & & & & \object{DENIS 29} \\
 \object{ChaI~607} & 11 08 12.1 & -77 18 54.0 &  22.8: &  19.9; &  20.4; &  23.2: &  22.6; &  L0 (M0.5) & star  & \object{KG36}, \object{ISO 130}, \\
& & & & & & & & & & \object{DENIS-P J1108.2-7718} \\
 \object{ChaI~608}$^{\mathrm{g}}$ & 11 08 00.6 & -77 17 31.2 & 17.79 & 15.51 & 17.82 & 19.26 & 18.72 & M4 & star & \object{CHXR 30}, \object{KG 30}, \\
& & & & & & & & & & \object{DENIS 58} \\
\hline
\end{tabular}
   \end{table*}

\newpage
   \begin{table*}[ht]
\centering
\scriptsize
\textbf{\footnotesize Table~\ref{tab:chaiphot}} (continued) \\        
\begin{tabular}{p{0.07\linewidth}c c c c c c c l l l l}
            \hline
            \hline
Name & $\alpha$ (2000) & $\delta$ (2000) & R & I & H$\alpha$ & M855 & M915 & SpT$^{\mathrm{c}}$ & Classification & Other Names$^{\mathrm{d}}$ \\
            \hline
 \object{ChaI~609} & 11 07 57.8 & -77 17 26.9 &  19.78 &  17.27 &  19.3; &  20.91 &  20.32 & M5 & star & \object{Baud 38}, \object{KG 26}, \\
& & & & & & & & & &  \object{DENIS 55} \\
 \object{ChaI~610} & 11 08 00.5 & -77 15 32.4 &  18.09 &  15.49 &  17.89 &  19.34 &  18.47 & M7 & BD &  \object{KG 29}, \object{ISO 121} \\
 \object{ChaI~613} & 11 08 27.1 & -77 15 55.8 &  19.38 &  16.73 &  18.84 &  20.60 &  19.62 & M8 & BD &  \object{ISO 147}\\
 \object{ChaI~615} & 11 09 05.7 & -77 09 59.0 &  16.02 &  14.09 &  15.76 &  17.99 &  17.45 & M4 & star & \object{Hn7}, \object{ISO 174}, \\
& & & & & & & & & & \object{DENIS 71} \\
 \object{ChaI~616}$^{\mathrm{e}}$ & 11 09 58.8 & -76 59 15.5 & 14.78 & 13.22 & 14.76 & 17.11 & 16.78 & $<$M4 & star & \object{KG 113}, \object{ISO 229}  \\
 \object{ChaI~618} & 11 06 50.4 & -76 59 29.1 &  16.97 &  15.03 &  16.86 &  20.15 &  20.00 & $<$M4 & star & \\
 \object{ChaI~619} & 11 04 37.2 & -77 02 14.5 &  13.26 &  11.89 &  13.12 &  15.77 &  15.53 & $<$M4 & star &  \\
 \object{ChaI~622} & 11 05 15.2 & -77 11 29.7 &  15.10 &  13.26 &  14.82 &  17.09 &  16.65 & M4 & star & \object{Hn4}, \object{DENIS 30}  \\
 \object{ChaI~625} & 11 10 29.1 & -77 17 00.4 &  16.18 &  14.04 &  15.89 &  17.92 &  17.21 & M6 & trans. obj. & \object{Hn12}, \object{DENIS 94}  \\
 \object{ChaI~629} & 11 11 23.3 & -77 05 54.7 &  15.87 &  13.95 &  15.64 &  17.80 &  17.36 & $<$M4 & star & \\
 \object{ChaI~708} & 11 09 13.9 & -76 28 40.1 &  15.89 &  13.99 &  15.61 &  18.03 &  17.43 & M5 & star & \object{Hn 8}, \object{KG 75}, \\
& & & & & & & & & & \object{DENIS 73} \\
 \object{ChaI~709} & 11 08 51.0 & -76 25 13.8 &  16.81 &  14.79 &  16.26 &  18.87 &  18.20 & M5.5 & star & \object{T37}, \object{Sz28}, \object{KG 54},  \\
& & & & & & & & & & \object{ISO 157}, \object{DENIS 68}  \\
 \object{ChaI~710} & 11 10 22.3 & -76 25 14.2 &  19.09 &  16.54 &  18.62 &  20.66 &  19.50 & M9 & BD &  \\
 \object{ChaI~711} & 11 07 11.9 & -76 25 50.2 &  18.37 &  16.15 &  18.06 &  20.01 &  19.32 & M5.5 & star &  \\
 \object{ChaI~712} & 11 06 32.8 & -76 25 21.0 &  19.42 &  16.96 &  19.1; &  20.93 &  19.92 & M8 & BD &  \\
 \object{ChaI~717} & 11 08 02.5 & -76 40 34.3 &  17.92 &  15.54 &  17.68 &  19.58 &  18.68 & M7 & BD &  \\
 \object{ChaI~721} & 11 09 51.6 & -76 45 45.4 &  16.06 &  14.30 &  15.92 &  18.09 &  17.78 & $<$M4 & star & \object{ISO 214} \\
 \object{ChaI~726} & 11 09 52.2 & -76 39 13.0 &  19.59 &  16.96 &  18.45 &  20.87 &  19.92 & M7.5 & BD &  \object{GK 29}, \object{KG 106}, \\
& & & & & & & & & & \object{ISO 217} \\
 \object{ChaI~731} & 11 09 22.8 & -76 34 32.2 &  19.94 &  17.55 &  18.88 &  21.11 &  20.74 & $<$M4 & star & \object{C1-6}, \object{NIR 10}, \object{ISO 189} \\
 \object{ChaI~735} & 11 08 55.1 & -76 32 40.1 &  18.03 &  15.76 &  17.12 &  19.63 &  18.95 & M5.5 & star & \object{C1-25}, \object{NIR 19}, \object{KG 59}, \\
& & & & & & & & & & \object{ISO 165}  \\
 \object{ChaI~737} & 11 08 51.9 & -76 32 50.3 &  20.3; &  17.61 &  20.0; &  21.60 &  20.62 & M8 & BD  &  \\
 \object{ChaI~742} & 11 12 03.4 & -76 37 03.7 &  16.28 &  14.13 &  15.90 &  18.18 &  17.43 & M6 & trans. obj. & \object{Hn16}, \object{CHXR 84}, \\
& & & & & & & & & & \object{DENIS 105}  \\
 \object{ChaI~743} & 11 11 45.4 & -76 36 50.1 &  21.4: &  18.60 &  21.0: &  22.8; &  21.47 & M8 & BD &  \\
 \object{ChaI~747} & 11 13 24.6 & -76 29 23.2 &  15.29 &  13.69 &  14.94 &  17.67 &  17.32 & $<$M4 & star & \object{Hn18}, \object{DENIS 116}  \\
 \object{ChaI~748} & 11 13 29.8 & -76 29 01.7 &  15.23 &  13.51 &  14.94 &  17.48 &  17.06 & $<$M4 & star & \object{Hn19}, \object{DENIS 118} \\
 \object{ChaI~749} & 11 12 59.4 & -76 16 53.6 &  13.57 &  11.93 &  13.23 &  15.98 &  15.63 & $<$M4 & star &  \\
 \object{ChaI~751} & 11 06 42.0 & -76 35 49.0 &  15.28 &  13.63 &  14.45 &  17.61 &  17.13 & M4 & star &  \\
 \object{ChaI~752} & 11 08 40.8 & -76 36 07.8 &  14.15 &  12.68 &  13.77 &  16.65 &  16.38 & $<$M4 (M2) & star & \object{CHX 13a}, \object{KG 52}, \\
& & & & & & & & & & \object{ISO 153}, \object{DENIS 65}  \\
 \object{ChaI~753} & 11 09 53.5 & -76 34 25.5 &  15.10 &  13.52 &  13.87 &  17.31 &  17.09 & $<$M4 & star &  \\
 \object{ChaI~754} & 11 12 48.8 & -76 47 06.7 &  15.46 &  13.79 &  15.15 &  17.78 &  17.34 & $<$M4 & star &  \\
 \object{ChaI~755} & 11 10 56.2 & -76 45 32.7 &  16.28 &  13.93 &  15.67 &  17.94 &  17.02 & M7 & BD &  \\
 \object{ChaI~757} & 11 12 10.0 & -76 34 36.8 &  15.00 &  13.12 &  14.51 &  17.18 &  16.60 & M4 & star &  \\
 \object{ChaI~759} & 11 13 33.7 & -76 35 37.8 &  15.37 &  13.59 &  15.08 &  17.57 &  17.11 & M4 & star &  \\
 \object{ChaI~760} & 11 10 11.5 & -76 35 29.3 &  15.56 &  13.89 &  15.29 &  17.74 &  17.47 & $<$M4 (M1.5) & star  & \object{NIR 45}, \object{KG 121}, \\
& & & & & & & & & & \object{ISO 237}, \\
& & & & & & & & & & \object{DENIS-P J1110.2-7635}  \\
\hline       
\end{tabular}
\begin{list}{}{}
\item[$^{\mathrm{a}}$] H$\alpha$, M855 and M915 magnitudes are instrumental
magnitudes (see text).
\item[$^{\mathrm{b}}$] Photometric errors: blank: 0.05; semicolon: 
                        0.1~mag; colon: 0.2~mag 
\item[$^{\mathrm{c}}$] The estimated errors in the spectral types are
                        1 subclass in the range M6-M9 and 2 subclasses in the
                        range M4-M6 (see text). Spectral types from the 
                        literature are indicated in parenthesis.
\item[$^{\mathrm{d}}$] References: Sz\#: Schwartz (\cite{schwartz77}); Hn\#:
        Hartigan (\cite{hartigan92}); T\#: Schwartz (\cite{schwartz91}); 
        CHXR\#: Feigelson et al. (\cite{feigelson93}); C\#: Prusti et al. 
        (\cite{prusti91}); NIR\#: Oasa et al. (\cite{oasa99}); Baud\#: Baud
        et  al. (\cite{baud84}); CHX\#: Feigelson \& Kriss \cite{feigelson89};
        B\#: Lawson et al. (\cite{lawson96}); ISO\#: Persi et al.
        (\cite{persi00}), P. Persi, priv. communication; DENIS\#: Cambr\'esy 
et
        al. (\cite{cambresy98}, Table 1); DENIS-P\#: Cambr\'esy et al.
        (\cite{cambresy98}, Table 2); GK\#: G\'omez \& Kenyon (\cite{gomez01});
        KG\#: Kenyon \& G\'omez (\cite{kenyon01}).
\item[$^{\mathrm{e}}$] This object does not have H$\alpha$ emission according 
        to its (H$\alpha$--R) colour.
\item[$^{\mathrm{f}}$] Ambiguous counterpart (see text).
\item[$^{\mathrm{g}}$] The (R--I) colour of this object is slightly too blue 
        according to our selection criteria.
\end{list}
   \end{table*}

%
\subsection{Spectral Types}\label{sec:spt}

\subsubsection{A Photometric Classification}\label{sec:calib}

        To confirm the membership of our candidates to the Chamaeleon~I cloud,
we should determine their spectral type. Since there are many objects spread
over a large region, and therefore spectroscopy is not easy, we attempt to
identify late-type objects exclusively from photometric observations.

        M-dwarfs are usually classified by means of their TiO and VO 
absorption features, that get deeper with later M-subspectral type 
(Kirkpatrick et al. \cite{kirkpatrick91}; Mart\'{\i}n et al. \cite{martin99}),
but disappear in K- and L-type objects. While TiO bands begin to saturate for
objects of spectral type later than M5, VO features  become very prominent for
late M-type objects. The WFI has a medium-band filter, M855, covering some
important TiO and VO absorption features that get deeper with later spectral
type in M-objects (see Fig.~5 in Kirkpatrick et al. \cite{kirkpatrick91}).
Another medium-band filter, M915, lies in a wavelength range which is
relatively featureless in late-type  objects. For the calibration, several
objects with known spectral type from K7 to L3 were observed, and a correlation
was established between the spectral type and the (M855--M915) colour index.

        We selected our spectral calibrators mostly from the Gliese-Jahreiss 
catalogue (H. Jahreiss, private communication). Suitable objects must be 
observable from the southern hemisphere, not form multiple systems, and not be
variable. Unfortunately,  the catalogue does not contain many stars of the
latest spectral types  (M8-L3), and very few fulfilled the required conditions.
For this reason,  we also selected some objects from the LHS catalogue (Luyten
\cite{luyten79}) and from the surveys DENIS (Delfosse et al.
\cite{delfosse99}; Mart\'{\i}n et al. \cite{martin99}) and 2MASS (Kirkpatrick
et al. \cite{kirkpatrick00}), although in these cases we could not be sure that
the objects were  single-star systems. 

        A first set of calibrators was observed in two nights (on the 30th and
31st of May 1999) of the observing run in which our Chamaeleon~I fields were
taken. Once the feasibility of our method was demonstrated, a second  and
larger set was observed on the 23rd December 2001 to improve our initial
calibration. In total, we observed 22 objects in the M855 and M915 filters, but
three of them  were rejected because they seemed double, or because of 
technical problems during the observations. Data reduction and aperture 
photometry were performed in the way explained in Sect.\ref{sec:red}. 
Table~\ref{tab:mcalphot} summarizes our photometry results and other relevant 
data of our final calibrator list. We recall that no absolute photometric 
system is available for the M855 and M915 passbands. We could not use these 
objects to establish an absolute magnitude scale in these filters because for 
most of them the distance is still unknown.

%
   \begin{table*}
     \centering
     \scriptsize
      \caption[]{\footnotesize Our sample of spectral standards\,$^{\mathrm{ab}}$}
         \label{tab:mcalphot}
     $$ 
         \begin{tabular}{p{0.19\linewidth}l c c l l l l l}
            \hline
            \hline
Name$^{\mathrm{c}}$ & $\alpha$ (2000) & $\delta$ (2000) & M855 & M915 & SpT\\
            \hline
 \object{GJ 13460}               & 09 28 51 &  -09 16 00 &  13.12 &  13.19 & K7
\\
 \object{GJ 7 59}                & 06 51 42 &  -43 53 12 &  12.96 &  13.02 & M0
\\ 
 \object{GJ 494}                 & 13 00 47 & \,12 22 33 &  11.83 &  11.69 &
M0.5 \\
 \object{GJ 7 58}                & 06 39 50 &  -61 28 48 &  12.71 &  12.77 & M1
\\
 \object{GJ 22066}               & 08 16 08 & \,01 18 12 &  12.24 &  12.26 & 
M2 \\
 \object{GJ 4 659}               & 07 38 41 &  -21 13 30 &  13.59 &  13.56 & M3
\\
 \object{GJ 402}                 & 10 50 52 & \,06 48 29 &  12.97 &  12.64 & M3
\\
 \object{GJ 4 512}               & 05 36 00 &  -07 39 00 &  14.40 &  14.20 & M4
\\
 \object{GJ 21055}               & 03 09 00 & \,10 01 24 &  15.99 &  15.69 & M5
\\
 \object{GJ 4 195}               & 02 02 16 & \,10 20 18 &  16.19 &  15.73 & M6
\\
 \object{LHS 3003}               & 14 56 38 &  -28 09 50 &  16.57 &  15.72 & M7
\\
 \object{GJ 5 14}                & 02 52 26 & \,00 56 18 &  19.81 &  19.27 & M8
\\
 \object{LHS 2065}               & 08 53 36 &  -03 29 32 &  18.48 &  17.40 & M9
\\
 \object{GJ 4 619}               & 07 07 53 &  -49 00 48 &  20.22 &  19.42 & M9
\\
 \object{DENIS-P J0909.9-0658}   & 09 09 57 &  -06 58 06 &  20.67 &  20.19 & 
L0 \\
 \object{2MASSW J0832045-012835} & 08 32 05 &  -01 28 35 &  20.98 &  20.57 & 
L1.5 \\
 \object{2MASS J1342236+175156}  & 13 42 36 & \,17 51 56 &  19.81 &  19.80 & 
L2 \\
 \object{2MASSW J0928397-16031}  & 09 28 40 &  -16 03 12 &  21.30 &  21.34 & 
L2 \\
 \object{DENIS-P J1058.7-1548}   & 10 58 47 &  -15 48 00 &  21.18 &  20.71 & 
L2.5 \\
 \object{DENIS-P J1047.5-1815}   & 10 47 31 &  -18 15 58 &  21.40 &  20.60 & 
L2.5 \\
            \hline
         \end{tabular}
     $$ 
\begin{list}{}{}
\item[$^{\mathrm{a}}$] M855 and M915 magnitudes are instrumental magnitudes.
\item[$^{\mathrm{b}}$] Photometric errors: blank: 0.05; semicolon: 
                        0.1~mag; colon: 0.2~mag 
\item[$^{\mathrm{c}}$] References: GJ\#: H. Jahreiss (private communication); 
LHS\#: Luyten (\cite{luyten79}); DENIS: Delfosse et al. (\cite{delfosse99}), 
Mart\'{\i}n et al. (\cite{martin99}); 2MASS: Kirkpatrick et al. 
(\cite{kirkpatrick00}).
\end{list}
   \end{table*}
%
%
   \begin{figure}
   \centering
    \includegraphics[width=6cm, angle=-90, bb= 50 70 550 775]{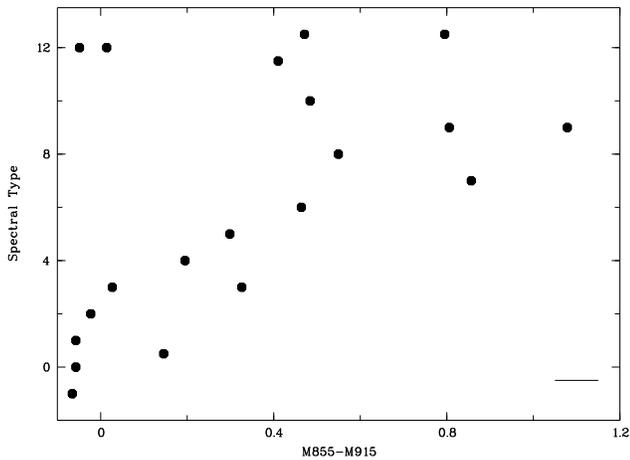} \hfill
      \caption{Spectral type versus (M855--M915) colour index
                for our set of spectral calibrators. Numbers from 0 to 9
                correspond to the M subspectral classes, and from 10 on to the
                L spectral sequence. For mid- to late M-type objects (M4-M9), a
                correlation between colour and spectral type is clearly seen.
                The bar at the lower right corner of the plot shows the average
                error in the (M855--M915) colours. 
               }
         \label{fig:mcal}
    \end{figure}

        For mid- to late M-type objects (M4-M9), a correlation between colour 
and spectral type is clearly found (see Fig.~\ref{fig:mcal}): Objects with 
later spectral type also have a higher (M855--M915) colour index. This is not 
seen for early M-type objects, because the features considered are not 
prominent in their spectra. The observed scatter probably is due mainly to
differences in age (see also Sect.~\ref{sec:young});
metallicity differences may also play a role. The best fit obtained in this 
range is:

\begin{equation}\label{eq:fit1}
SpT=(6.2\pm1.2)\cdot(\mathrm{M}855-\mathrm{M}915)+(2.8\pm0.7)
\end{equation}

\noindent
where $SpT$ is an integer (4-9) denoting the M subspectral class. The quoted 
errors are the standard deviations of the corresponding coefficients.

For early L-type objects (L0-L2), an  anticorrelation is observed, as expected 
from the progressive weakening of the TiO and VO features. However, this 
anticorrelation seems to break down for objects of spectral type later than L2.
This is probably due to the appearance of the alkali absorption features, a 
characteristic of the L spectral sequence. One  Cs~I line lies towards the 
centre of the wavelength range covered by our M855 filter, and it starts to be 
important for L3 and later-type objects (see Kirkpatrick et al. 
\cite{kirkpatrick99}, their Figs.~4, 6 and 7). Since L3 is the limit of our 
observed spectral range, we cannot state whether a sequence similar to that of 
late M-objects, but based on the alkali absorption features, could be 
established also for late L-objects. In the range M9-L2, we obtain as best fit:

\begin{equation}\label{eq:fit2}
SpT=(-3.1\pm0.6)\cdot(\mathrm{M}855-\mathrm{M}915)+(12.0\pm0.4)
\end{equation}

\noindent
where $SpT$ is now in the range 9-12 (a value of 10 corresponds to a spectral 
type L0).

%
   \begin{figure}
   \centering
    \includegraphics[width=6cm, angle=-90, bb= 50 70 550 775]{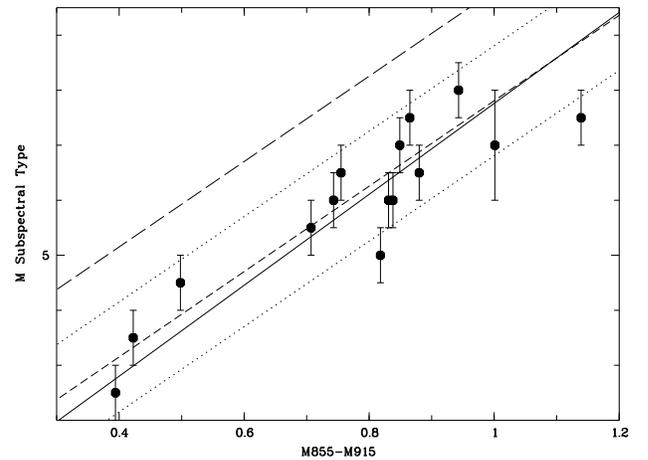} 
\hfill
      \caption{Spectral type versus (M855--M915) colour for the objects in 
                Chamaeleon~I with known spectral type. The errors are taken
                from \cite{comeron00}. A correlation between colour and
                spectral type is clearly seen. The long-dashed line is the fit
                obtained with our spectral calibrators. The solid line is the
                fit for this second set of objects. Both lines are almost
                parallel, as indicated by the short-dashed line, which
                corresponds to the fit on the old objects, but shifted 2
                subclasses. The dotted lines indicate the estimated error of
                about 1 subclass.}
              \label{fig:sptnc}
    \end{figure}
%

\subsubsection{Application to Young Objects}\label{sec:young}

        In a next step, we used Eq.~\ref{eq:fit1} to derive spectral types for
the known objects in  Chamaeleon~I. 

The results are shown in Fig.~\ref{fig:sptnc}. The young objects appear to 
have, according to our fit, later spectral types than the ones derived from 
spectroscopic observations. However, the trend to increasing colour with later 
spectral type is clearly reproduced. The solid line shows a fit performed on 
this set of young objects. This line is almost parallel to the one obtained 
from the fit to the (old) field objects, just being shifted to earlier spectral
types by about 2 subclasses.

        We interpret this shift as a consequence of the lower value of $\log g$
for these young objects. Work by several authors (e.g. Mart\'{\i}n et al. 
\cite{martin96}; Luhman et al. \cite{luhman97}) shows that young M-type objects
(characterized by low-gravity atmospheres) display stronger TiO and VO 
absorption than field dwarfs of the same spectral class (high-gravity 
atmospheres). Stronger TiO and VO absorption implies fainter M855 magnitudes 
and redder (M855--M915) colours. The age difference should also be responsible,
to some extent, for the scatter of our calibrators. 

\begin{figure*}
   \centering
    \includegraphics[width=6cm, angle=-90, bb= 50 70 550 775]{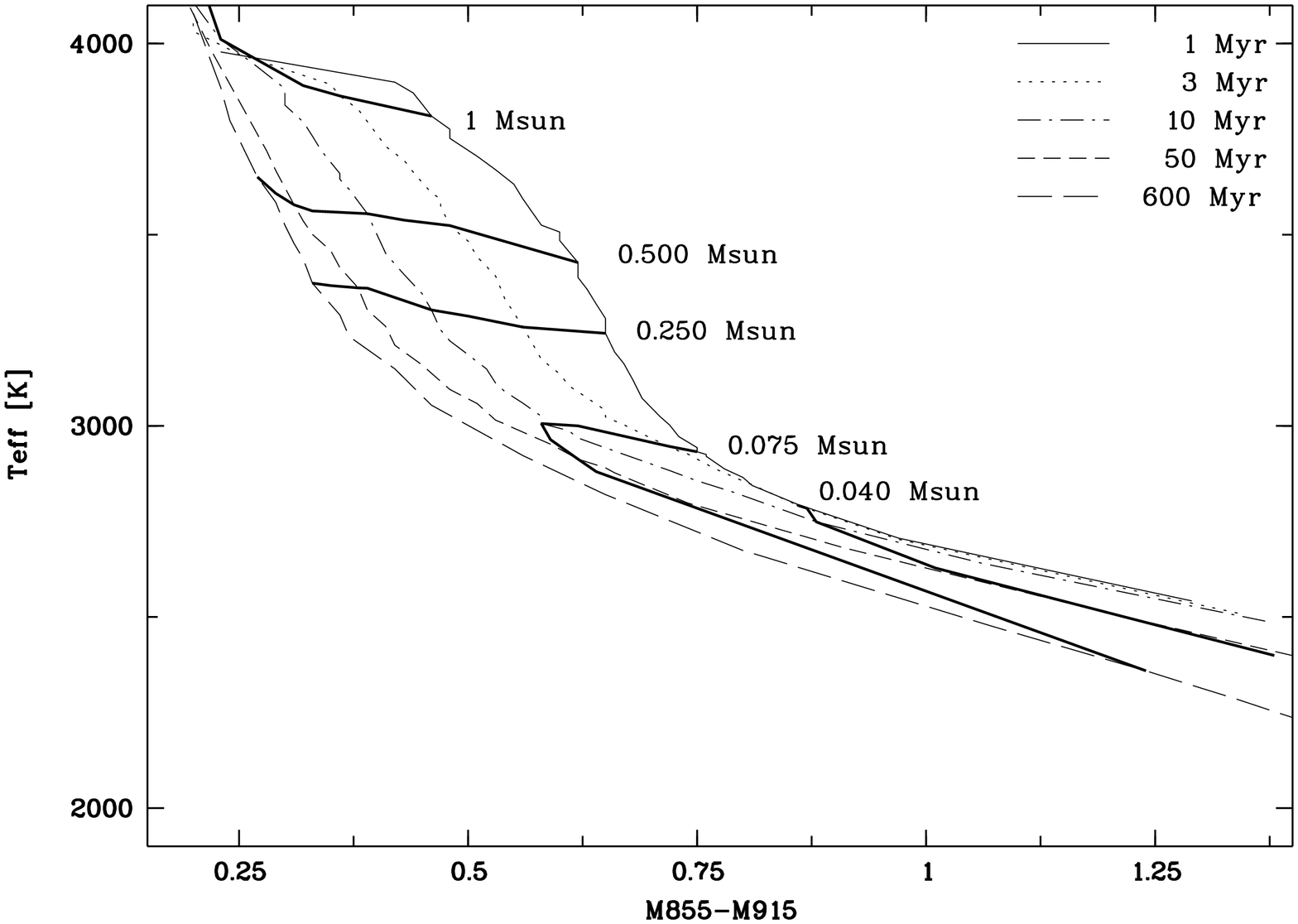} \hfill
    \includegraphics[width=6cm, angle=-90, bb= 50 70 550 775]{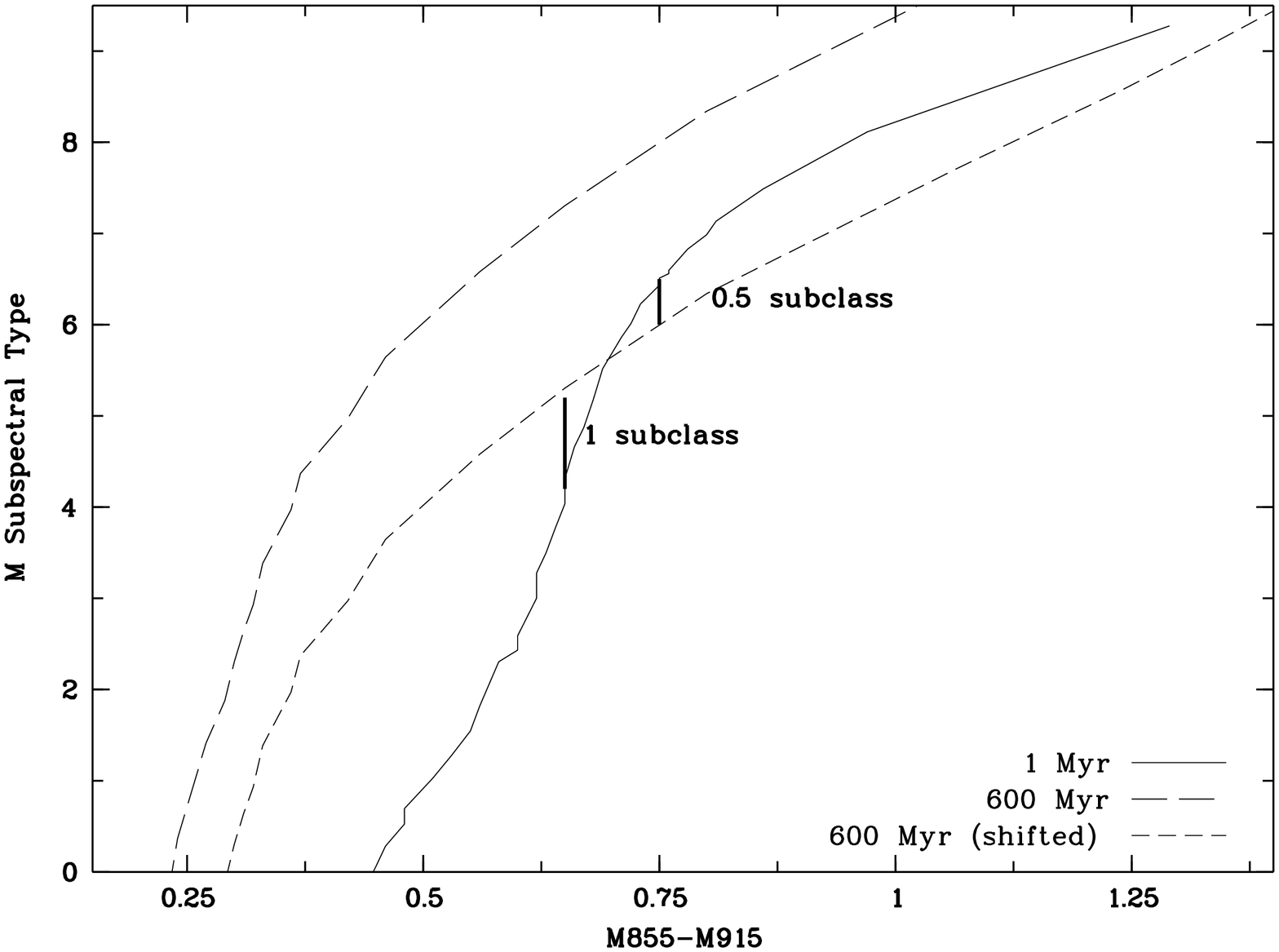}    
      \caption{\emph{Left panel:} Isochrones of effective temperature versus 
                theoretical (M855--M915) colour index from the Baraffe et al. 
                (\cite{baraffe98}) models. The curves of equal mass are also 
                plotted. For the same value of the colour index, older objects
                have a lower temperature, thus a later spectral type. Hence, by
                using a fit to old objects, the temperatures  and masses of
                young very low-mass objects are overestimated. 
                \emph{Right panel:} Spectral type versus theoretical 
                (M855--M915) colour index for the 1~Myr and the  600~Myr
                isochrones. At low values of the colour index, older objects
                are several subclasses later than very young objects. This
                difference decreases to at most 1 subclass for objects later
                than M4 if the oldest isochrone is shifted by 2 subclasses.}  
                \label{fig:sptage}
    \end{figure*}

The effect of age (i.e. gravity) in our medium-band photometry can be tested
with the help of theoretical models. The left panel of Fig.~\ref{fig:sptage}
shows different isochrones in a (T$_{eff}$, M855--M915) diagram from the
non-grey models of Baraffe et al. (1998; I. Baraffe, private communication).
The represented ages range from 1 to 600~Myr. All curves tend to converge for
effective temperatures higher than 4000~K, corresponding to late K-type
objects. For lower temperatures, the younger objects tend to exhibit redder
(M855--M915) colours than the older ones of same spectral type. The biggest
differences are observed in the temperature range between 4000 and 2800~K,
roughly corresponding to spectral types from K7 to M6. Here, a 1~Myr old object
will have the same colour as a 600~Myr old M-dwarf that is about 800~K cooler.
Even compared to a 3~Myr old object a remarkable temperature difference is
predicted  ($\sim 200$~K). However, for cooler objects (T$_{eff}\lesssim
2800$~K), the colour difference is no longer significant at very young ages
(less than 10~Myr), and the difference between young and old objects is notably
decreased (from 800 to about 150~K). On the other hand, the isochrones for 
intermediate ages (10-70~Myr) appear more or less parallel through the whole 
represented range. The oldest isochrone (600~Myr) only diverges from them at
very low temperatures. 

        The relation between spectral type and (M855--M915) colour index can be
better understood with the help of the right panel of Fig.~\ref{fig:sptage}.
Here we have adopted the T$_{eff}$ scale versus spectral type for pre-main
sequence objects from Luhman (\cite{luhman99}):

\begin{center}
\begin{equation}
\mathrm{T}_{eff}=3850-141.0 \cdot SpT
\end{equation}
\end{center}

\noindent 
It is immediately seen that the behaviour of the curves is not exactly linear: 
For blue (M855--M915) colours (earlier spectral types) the difference between 
old and young objects can be as high as 5 subclasses, but this difference is
reduced towards redder colours (later spectral types). If we shift the oldest
isochrone 2 subclasses, however, the difference is not greater than 1 subclass
for spectral types later than M4. The value of this shift is the approximated
offset that we find between our fit to (old) field M-dwarfs and the (young)
objects in Chamaeleon~I. We therefore conclude that the observed offset between
our initial spectral type calibration and the reported spectral types for the
objects in Chamaeleon~I is mainly due to the difference in age between them and
our calibrators, which implies a different value of $\log g$. To account for
this effect, we shifted our derived scale 2 subclasses towards earlier 
spectral types (see Fig.~\ref{fig:sptnc}). In most of the cases, the so derived
spectral types coincide with the  ones in \cite{comeron99} and \cite{comeron00}
(obtained from object spectra) with an error of one subclass at most (see
Table~\ref{tab:nc}). For  early M-type objects, the error is certainly larger,
due to the saturation of the (M855--M915) colour. We conservatively consider
an error of two subclasses in the range M4-M6.

         We cannot test our (M855--M915) colour index for L-dwarfs in the same
way, because no L-type objects are yet known in Chamaeleon~I. The model curves
of Fig.~\ref{fig:sptage} do not go deep enough into the L-type temperature
range to test the observed behaviour. Therefore, we can only tentatively
assign L0 to L2 spectral types to the objects found at the bottom left part of
a (M855, M855--M915) colour-magnitude diagram. We note, however, that highly 
extincted M-type objects would also fall in the same region of the diagram, 
making this classification somewhat uncertain. Hence, it is difficult, at the 
present stage, to distinguish between a L-type object and a rather extincted 
M-type object with this method (see also the discussion in the following 
section). A more detailed analysis is required before a reliable determination 
of the L subspectral class can be achieved with this photometric technique.

%
%
%
\subsubsection{Spectral Types of the Brown Dwarf Candidates} 
\label{sec:sptchai}
 
\begin{figure}
   \centering
   \includegraphics[width=6cm, angle=-90, bb= 50 70 550 775]{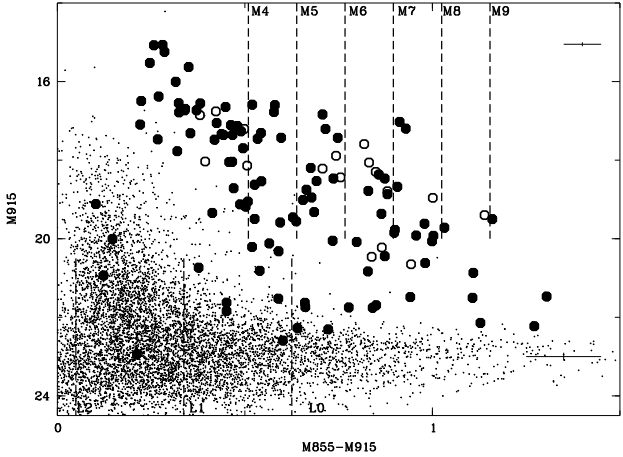} \hfill
     \caption{(M915, M855--M915) colour-magnitude diagram for our surveyed 
                field in Chamaeleon~I. Symbols as in Fig.~\ref{fig:richai}. A
                scale for the identification of the spectral type is also
                indicated (see text).}
         \label{fig:mmchai}
    \end{figure}

        Fig.~\ref{fig:mmchai} shows the (M915, M855--M915) colour-magnitude 
diagrams for our Chamaeleon~I field. Our derived scale for the spectral  type
identification is also shown. Clearly, many of our candidates have mid to late
spectral type M. About one fifth of them have spectral type M7 or later. Given
that the stellar/substellar boundary at the age of Chamaeleon~I is set at a
spectral type of about M6 by the theoretical models, we classify these objects
as brown dwarfs. There is a smaller group of seven or eight objects with
spectral types between M5.5 and M7, in the transition region from low-mass
stars to brown dwarfs. Around 50 objects have spectral types M5 or earlier, so
they are very low-mass stars. For the faintest objects, a turning back towards
lower values of the (M855--M915) colour is seen in Fig.~\ref{fig:mmchai} at
spectral type M9, which may be indicative of the beginning of the L-dwarf 
sequence. If their belonging to the cloud can be confirmed, these objects
might be low-mass brown dwarfs. 

        The photometry results and the derived spectral type for all our 
candidates are summarized in Table~\ref{tab:chaiphot}. The values obtained from
Eqs.~\ref{eq:fit1} and \ref{eq:fit2} have been rounded to the closest value in 
steps of 0.5 subclasses. Although 23 objects from our survey list had already 
been proposed as cloud members in previous optical and/or X-ray studies of 
Chamaeleon~I (see references in Table~\ref{tab:chaiphot}), many of them still 
lacked a spectral type determination. We have checked our photometrically 
derived spectral types with those given by Lawson et al. (\cite{lawson96}) and 
Gauvin \& Strom (\cite{gauvin92}) for the objects in common with these authors 
(7). In general there is good agreement between their and our spectral types, 
except for the star Sz~10 (ChaI~456), to which Gauvin \& Strom assign a 
spectral type of M0.5, while our photometric classification yields a spectral 
type of M5. We do not see, however, any other object near ChaI~456 that could 
be the one observed by Gauvin \& Strom. DENIS photometry is also available for 
this object (Cambr\'esy et al. \cite{cambresy98}, see Sect.~\ref{sec:ir} 
below). The I magnitude measured by these authors is consistent with our own 
result within the estimated errors, while their $JKs$ photometry agrees well 
with that given by Gauvin \& Strom (\cite{gauvin92}). We are thus quite 
confident that our identification of ChaI~456 with Sz~10 is correct. However, 
the measured I magnitude (13.40~mag) appears too bright for an M5 object. The 
discrepancy could be explained if Sz~10 were an unresolved binary. In that 
case, Gauvin \& Strom would have observed the spectrum of the primary. We would
wrongly assign a later spectral type to the unresolved pair, because our 
spectral type calibration is only based on the (M855--M915) colour index (which
should be redder for a binary object), and not on magnitudes (which should be 
brighter). As a matter of fact, in Table~\ref{tab:nc} we see that CHXR22, a 
known binary formed by a K7 and a M2 object that appear unresolved in our 
images, has an estimated spectral type of M4 according to our classification. 
Thus an hypothetical secondary of Sz~10 would probably have a spectral type 
earlier than M5, definitively being a star.

        More recently, G\'omez \& Persi (\cite{gomez02}) presented infrared
spectroscopic observations of several mid-infrared sources from a previous
ISOCAM survey in Chamaeleon~I (Persi et al. \cite{persi00}). Six of their
objects are also included in our list of probable cloud members (some of them 
in common with other authors). They give an error of 2~subclasses for their 
spectral type classification. While for four of the objects the agreement 
between their and our spectral types is rather good, for the other two we find 
again a large discrepancy: ISO~252 (our object ChaI~406) is classified as an M2
object according to these authors, but we derive a spectral type of M8 for it. 
Its I magnitude according to our and the DENIS measurement (around 17.2~mag) is
slightly too faint compared to that of other objects of similar spectral type 
in the cloud. On the other hand, the object ISO~250 (ChaI~405), which lies not 
too far from ISO~252, has a spectral type of M6 according to G\'omez \& Persi 
and of M5.5 according to our own classification, in spite of being brighter 
than ISO~252 in the I band. The discrepancy between both spectral type 
classifications for ISO~252 might be due to a locally higher extinction towards
this object. A similar situation is seen in the case of ISO~220 (ChaI~412), 
which we classify as M9 and G\'omez \& Persi as M4.5. The I magnitude for this 
object (greater than 18~mag according to both DENIS and our photometry) is very
faint compared to other objects of spectral type M4-M5 in the cloud (typically 
between 16 and 17~mag). The (R--I) colours of both ChaI~406 and ChaI~412 are 
too red for early M-type objects, which might indeed hint to the presence of 
moderate extinction.

We note, however, that the discrepancy between luminosity and spectral type
might also be due to the presence of circumstellar dust, as proposed by
Fern\'andez \& Comer\'on (\cite{fernandez01}) for the object LS-RCrA~1 in
Corona Australis. This object appears too faint compared to other objects of 
similar spectral type and age. More such subluminous objects have been
identified in other star forming regions.

The effect of extinction is more clearly seen when comparing our spectral types
with those  given by G\'omez \& Mardones (\cite{gomez03}). These authors
performed  near-infrared spectroscopy of 46 candidate young stellar objects
from several  photometric surveys in Chamaeleon~I. We have six objects in
common with this  work.\footnote{G\'omez \& Mardones (\cite{gomez03}) actually
observed seven  of our objects, but the spectrum of one of them, ChaI~427, was
too noisy to allow  a spectral type classification.} For three of them
(ChaI~403, ChaI~432 and ChaI~760), the spectral  classification given by
G\'omez \& Mardones agrees with ours within their and  our estimated errors.
This is not the case, however, for the other three  objects (ChaI~420, ChaI~428
and ChaI~607), whose spectral types are remarkably  warmer according to the
spectroscopic data. While we assign them spectral  types between M9 and L0 with
our photometric method, G\'omez \& Mardones classify  them as early M-type
objects (M0.5-M1). These authors obtain values between 2  and 5~mag for the
extinction A$_J$ in the $J$ band towards these objects,  which are among the
faintest ones in our survey. 

        These cases illustrate the limitations of our photometric spectral type
determination: We will estimate too late spectral types for some objects that
lie in regions of locally high extinction, which could then be wrongly
classified as brown dwarfs. The uncertainties are especially large for the 
faintest candidates in our sample, that could be identified as L-type objects 
in spite of being highly extincted early M-type stars.

On the other hand, if a particular object is a binary which appears unresolved 
in our WFI images, we are likely to assign a later spectral type to the 
combined pair. Nevertheless, among a total of 36 objects with known spectral 
type, we have only six clear discrepancies, meaning that in about 80\% of the 
cases our classification is reliable. This reliability is even higher (around 
90\%) for the brightest objects (I$<$19). On these objects, a linear fit of our
predicted spectral types versus the ones from the literature gives a RMS error 
of about one subclass, in agreement with our expectations.

        Out of 72 objects, we have classified 44 objects as low-mass 
stars (with spectral type earlier than M6), and 19 objects as brown dwarfs 
(with spectral type later than M6.5). We also find 6 transition objects with 
spectral types between M6 and M6.5, which will be considered brown dwarf 
candidates in the subsequent discussion. Three objects that could be 
low-mass brown dwarfs of spectral type L are also detected. However, in some 
cases the identification of our faint candidates as cloud members is not 
certain because of the large photometry errors in the lower part of the 
(H$\alpha$, H$\alpha$--R) colour-magnitude diagram and/or the lack of infrared 
excess (see Sect.~\ref{sec:ir}). Moreover, as explained in this section, their 
spectral type classification may be somewhat uncertain and requires 
spectroscopic confirmation. We will regard these objects as brown dwarf 
candidates, although keeping in mind that they might be embedded stars.

        In the subsequent analysis we do not include the doubtful objects 
ChaI~608 and ChaI~404. In the seven cases mentioned above where the 
photometric spectral type does not match with reported spectroscopic 
observations, the classification derived from object spectra was preferred.
Spectral types from the literature are indicated in parenthesis in the 
corresponding column of Table~\ref{tab:chaiphot}.

   \begin{table*}[ht]
  \scriptsize
  \centering
      \caption[]{\footnotesize Infrared photometry for candidate low-mass 
	        members of Chamaeleon~I$^\mathrm{a}$}
         \label{tab:irphot}
         \begin{tabular}{p{0.1\linewidth}l c c c c c c c c}
            \hline
            \hline
Name & $J_D$ & $Ks_D$ & $J$ & $H$ & $K$ & $L$ & F$_{6.7}$ (mJy) & F$_{14.3}$ 
(mJy) \\
            \hline
ChaI~403 & 13.28 & 10.83 &      &       &       &       &                 & \\
ChaI~404 & 12.34 & 10.80 &      &       &       &       &  20.0 $\pm$ 0.0 &  
5.0 $\pm$  1.2  \\
ChaI~405 & 12.34 & 10.80 &      &       &       &       &  20.0 $\pm$ 0.0 &  
5.0 $\pm$  1.2  \\
ChaI~406 & 13.75 & 12.30 &      &       &       &       &   5.5 $\pm$ 0.0 &  
6.0 $\pm$  1.2  \\
ChaI~410 & 14.07 & 13.02 &      &       &       &       &  15.3 $\pm$ 0.0 & 
17.6 $\pm$  0.3 \\
ChaI~411 & 12.65 & 11.06 & 12.74 & 11.65 & 11.10 & 10.67 &  15.6 $\pm$ 1.0 &  
16.5 $\pm$  1.5 \\
ChaI~412$^{\mathrm{c}}$ & 14.36 & 12.38 & 14.57 & 13.33 & 12.48 &       &  
5.3$\pm$ 
0.0 &   6.0 $\pm$  1.1  \\
ChaI~417 & 12.38 &  9.93 & 12.74 & 11.12 & 10.30 &  8.58 &  34.4 $\pm$ 1.0 & 
35.0 $\pm$  1.4 \\
ChaI~418 & 12.19 & 10.42 & 12.48 & 11.12 & 10.48 &  9.55 &  14.9 $\pm$ 1.0 & 
12.3 $\pm$  2.3  \\
ChaI~420 & 13.10 &  8.83 & 12.96 & 10.28 &  8.90 &  7.74 &  13.5 $\pm$ 1.0 & 
36.2 $\pm$  1.4  \\
ChaI~422 & 13.05 & 11.91 &      &       &       &       &   3.2 $\pm$ 0.0 & \\
ChaI~425 & 11.26 &  9.96 &      &       &       &       &  11.4 $\pm$ 0.0 &  
3.7 $\pm$  1.0 \\
ChaI~427 & 13.40 & 10.01 &      &       &       &       &                 & \\
ChaI~428 & 13.75 & 10.50 &      &       &       &       &                 & \\
ChaI~429 &       &      & 13.62 & 12.47 & 11.83 & 10.72 &                 & \\
ChaI~432 & 12.72 & 10.08 &      &       &       &       &  12.6 $\pm$ 1.0 & \\
ChaI~433 & 12.78 & 10.51 &      &       &       &       &   7.0 $\pm$ 1.0 & \\
ChaI~434 & 12.39 & 10.93 & 11.92 & 10.92 & 10.47 & 10.03 &   7.4 $\pm$ 1.0 & 
\\
ChaI~438 & 13.01 & 11.56 &      &       &       &       &   7.6 $\pm$ 1.0 &  
8.0 $\pm$  1.5  \\
ChaI~441 & 14.55 & 11.69 &      &       &       &       &   3.4 $\pm$ 0.0 & \\
ChaI~448 & 11.30 & 10.14 &      &       &       &       &  11.2 $\pm$ 0.0 & \\
ChaI~449 & 11.29 & 10.09 &      &       &       &       &  13.2 $\pm$ 1.0 & \\
ChaI~450 & 10.94 &  9.27 &      &       &       &       &  20.7 $\pm$ 0.0 & \\
ChaI~453 & 11.23 & 10.16 &      &       &       &       &  15.0 $\pm$ 0.0 &  
7.5 $\pm$  1.9 \\
ChaI~456 & 11.41 & 10.36 &      &       &       &       &                 & \\
ChaI~457 & 11.78 & 10.61 &      &       &       &       &                 & \\
ChaI~603 & 12.10 & 10.27 &      &       &       &       &  23.1 $\pm$ 1.0 & 
14.4 $\pm$  2.1 \\
ChaI~607 & 14.54 & 10.48 & 14.46 & 11.72 & 10.47 &  9.92 &  14.3 $\pm$ 1.0 & 
\\
ChaI~608 & 11.74 &  9.04 & 11.71 &  9.86 &  9.08 &  8.45 &                & \\
ChaI~609 & 13.08 &  9.87 & 13.19 & 11.00 &  9.61 &  8.29 &                & \\
ChaI~610 & 12.31 & 11.13 & 12.46 & 11.57 & 11.10 & 10.38 &   4.8 $\pm$ 1.0 & 
\\
ChaI~613 & 13.73 & 12.34 &      &       &       &       &   4.3 $\pm$ 1.0 & \\
ChaI~615 & 11.79 & 10.77 &      &       &       &       &   5.9 $\pm$ 0.0 & \\
ChaI~616 & 10.46 &  8.73 & 10.59 &  9.38 &  8.94 &  8.51 &  26.6 $\pm$ 1.0 &   
6.0 $\pm$  1.2  \\
ChaI~622 & 10.93 &  9.46 &      &       &       &       &                 & \\
ChaI~625 & 11.60 & 10.64 &      &       &       &       &                 & \\
ChaI~708 & 11.64 & 10.67 & 11.81 & 11.34 & 10.79 & 10.28 &                & \\
ChaI~709 & 12.45 & 11.30 & 12.38 & 11.66 & 11.25 & 10.53 &   7.5 $\pm$ 1.0 & 
\\
ChaI~721 & 11.10 &  9.26 &      &       &       &       &  24.0 $\pm$ 1.0 & \\
ChaI~726 & 13.28 & 11.70 & 13.36 & 12.34 & 11.65 & 10.76 &  10.3 $\pm$ 2.0 & 
\\
ChaI~731 & 13.13 &  9.32 &      &       &       &       & 383.1 $\pm$ 5.0 &
611.3 $\pm$ 19.3  \\
ChaI~735 & 13.03 & 11.38 & 12.92 & 11.96 & 11.36 & 10.90 &  12.0 $\pm$ 1.0 & 
\\
ChaI~742 & 11.69 & 10.66 &      &       &       &       &                 & \\
ChaI~747 & 11.78 & 10.64 &      &       &       &       &                 & \\
ChaI~748 & 11.53 & 10.45 &      &       &       &       &                 & \\
ChaI~752 & 10.45 &  9.09 & 10.56 &  9.56 &  9.20 &  8.92 &  26.8 $\pm$ 2.0 & 
\\
ChaI~760 & 10.77 &  8.51 & 10.87 &  9.34 &  8.55 &  7.69 & 186.3 $\pm$ 6.0 & 
238.0 $\pm$  8.8  \\

\hline
\end{tabular}
\begin{list}{}{}
\item[$^{\mathrm{a}}$] $J_DKs_D$: DENIS photometry (Cambr\'esy et al. 
        \cite{cambresy98}); $JHKL$: Photometry from Kenyon \& G\'omez 
        (\cite{kenyon01}).
\item[$^{\mathrm{b}}$] Photometry from G\'omez \& Kenyon (\cite{gomez01})
\end{list}
   \end{table*}
%

%
\subsection{Infrared Detections}\label{sec:ir}

        We cross-correlated our Chamaeleon~I object list with published 
detections by ISOCAM (Persi et al. \cite{persi00}) and DENIS (Cambr\'esy et 
al. \cite{cambresy98}), as well as the surveys by G\'omez \&  Kenyon
(\cite{gomez01}) and Kenyon \& G\'omez (\cite{kenyon01}). In this way, we got
infrared photometry for about 67\% of the objects in our sample. Of them, 47
have a near-infrared counterpart and 33 a mid-infrared counterpart.

        Persi et al. (\cite{persi00}) provide a list of faint Chamaeleon~I
members with  mid-infrared excesses from ISOCAM observations. We found 19
objects in common  in both lists. Another 11 objects from our sample have been 
detected by ISOCAM in one or both bands (P. Persi, private communication). The
ISOCAM list also provides DENIS I$JKs$ photometry for all the detected 
sources. In addition, we recover 23 objects  with near-infrared excess from
DENIS photometry according to Cambr\'esy et al. (\cite{cambresy98}). G\'omez et
al.  (\cite{gomez01}) report a list of 118 candidate members of the
Chamaeleon~I cloud  with infrared excess from $JHK$ observations, but we find
only two objects in  common with these authors. Kenyon \& G\'omez
(\cite{kenyon01}) provide $JHKsL$  photometry for a sample of young stellar
candidates in Cha\,I, including many  from the previous surveys and some new
detections. This sample contains 16 of  our objects. In total, we find 33
objects with a reported near-infrared excess. For the ones observed by several
authors, the photometry results are consistent within the quoted errors. The
available infrared photometry for our objects, together with the
cross-identifications and references are summarized in Table~\ref{tab:irphot}.
Surprisingly, only 10 objects exhibit both near- and mid-infrared excesses.
This represents less than 25\% of the total infrared detections in our sample.

        A dubious cross-identification regards the objects ChaI~404 and
ChaI~405, which lie relatively close at a distance of about 9$\farcs$5. The
coordinates given by Persi et al. (\cite{persi00}) for the ISOCAM source
ISO~250 match well the  position of ChaI~404, the fainter of the two, but the
I-band photometry from DENIS most probably corresponds to ChaI~405, which is
about one magnitude brighter than ChaI~404. In our ``reddest'' filter, M915,
the brightness difference between both sources is reduced to about 0.8~mag. It
is possible that ChaI~404 is brighter in the infrared than its neighbour, thus
leading to an eventual identification error. However, ChaI~404 does not show
evident H$\alpha$ emission according to its H$\alpha$--R colour, so its
membership to the Chamaeleon~I cloud is doubtful according to our selection
criteria. Given the ISOCAM pointing errors, we cannot exclude that ChaI~404 is
indeed a background object, and that ChaI~405 is the true counterpart to the
mid-infrared source.

        The detection of infrared excesses indicates that the Chamaeleon~I
candidate members  must be young. However, it should be taken into account that
criteria for identifying  objects with infrared excess are not homogenous among
the different authors.  In particular, we note that the G\'omez \& Kenyon
survey has few  objects in common with the other two, in spite of their spatial
overlap. For a detailed discussion of the infrared properties, we refer to the 
quoted works. Nonetheless, something has to be said about the extinction 
towards our objects, as it may affect their properties, and in particular, 
according to the previous section, their spectral type classification. 

\begin{figure}
   \centering
   \includegraphics[width=6cm, angle=-90, bb= 50 70 550 775]{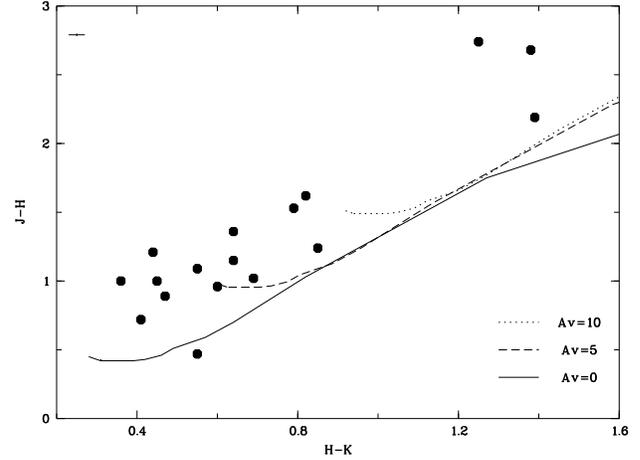} 
\hfill
     \caption{Colour-colour diagram for our candidates with near-infrared 
       photometry. The curves indicate the position of a 1~Myr isochrone from 
       Chabrier et al. (\cite{chabrier00}) for different extinction values. 
       The crossed lines on the upper left corner of the plot indicate the 
       average errors according to Kenyon \& G\'omez (\cite{kenyon01}). 
     }
         \label{fig:irdiag2}
    \end{figure}

In Fig.~\ref{fig:irdiag2} we show the near-infrared colour-colour diagram for
our candidates with $JHK$ photometry (G\'omez \& Kenyon \cite{gomez01}; Kenyon 
\& G\'omez \cite{kenyon01}). The solid curve is a 1~Myr isochrone from the 
Chabrier et al. (\cite{chabrier00}) models. The dashed and dotted lines 
indicate the position of this isochrone for extinction values of A$_V=5$~mag 
and  A$_V=10$~mag, respectively, computed with the law of Rieke \& Lebofsky 
(\cite{rieke85}). The colours of most of our candidates are consistent with 
extinction values between 0 and 10~mag. The three exceptions are ChaI~420, 
ChaI~607 and ChaI~609, that appear to be significantly reddened. For the two 
former objects, this result was expected after the discussion from 
Sect.~\ref{sec:sptchai}. Note that for the third object with a remarkably 
discrepant spectral type (ChaI~428) only DENIS photometry is available. 
ChaI~609, on the other hand, is very red for its estimated spectral type (M5).

Persi et al. (\cite{persi00}) used a ($J-K$, $K-$m(6.7)) colour-colour diagram 
to separate the effects from intrinsic infrared excess from those of 
reddening.

        The apparent magnitudes m(6.7) and m(14.3) corresponding to the ISOCAM 
fluxes (in Jy) are computed with the following relations:

\begin{center}
\begin{equation}\label{eq:isomag1}
\mathrm{m}(6.7)=-2.5\cdot \log\frac{\mathrm{F}_{\nu}(6.7\mu m)}{83.4}
\end{equation}
\begin{equation}\label{eq:isomag2}
\mathrm{m}(14.3)=-2.5\cdot \log\frac{\mathrm{F}_{\nu}(14.3\mu m)}{18.9}
\end{equation}
\end{center}

\begin{figure}
   \centering
   \includegraphics[width=6cm, angle=-90, bb= 50 70 550 775]{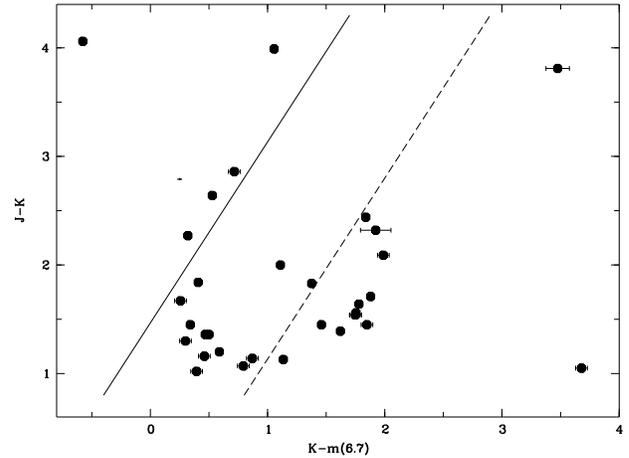} 
\hfill
     \caption{($J-K$, $K-$m(6.7)) colour-colour diagram for our candidates 
       detected in the near and mid-infrared. The objects with intrinsic 
       ($K-$m(6.7)) excess are found to the right of the assumed reddening 
       band, indicated by the two parallel lines.
     }
         \label{fig:irdiag1}
    \end{figure}

In Fig.~\ref{fig:irdiag1} we show the position of our objects in the  ($J-K$,
$K-$m(6.7)) diagram. The solid and dashed lines indicate the location of the 
reddening band computed by Persi et al. (2000). The objects with intrinsic
($K-$m(6.7))  excess are found to the right of the reddening band. ChaI~420 and
ChaI~607 are  clearly separated from the rest, having high values of the
($J-K$) colour  index,  but low or moderate ($K-$m(6.7)) colours. This diagram
confirms that these two  objects are heavily reddened, and that many from our
candidates suffer from  moderate extinction.

%
   \begin{figure}
   \centering
  \includegraphics[width=6cm, angle=-90, bb= 50 70 550 775]{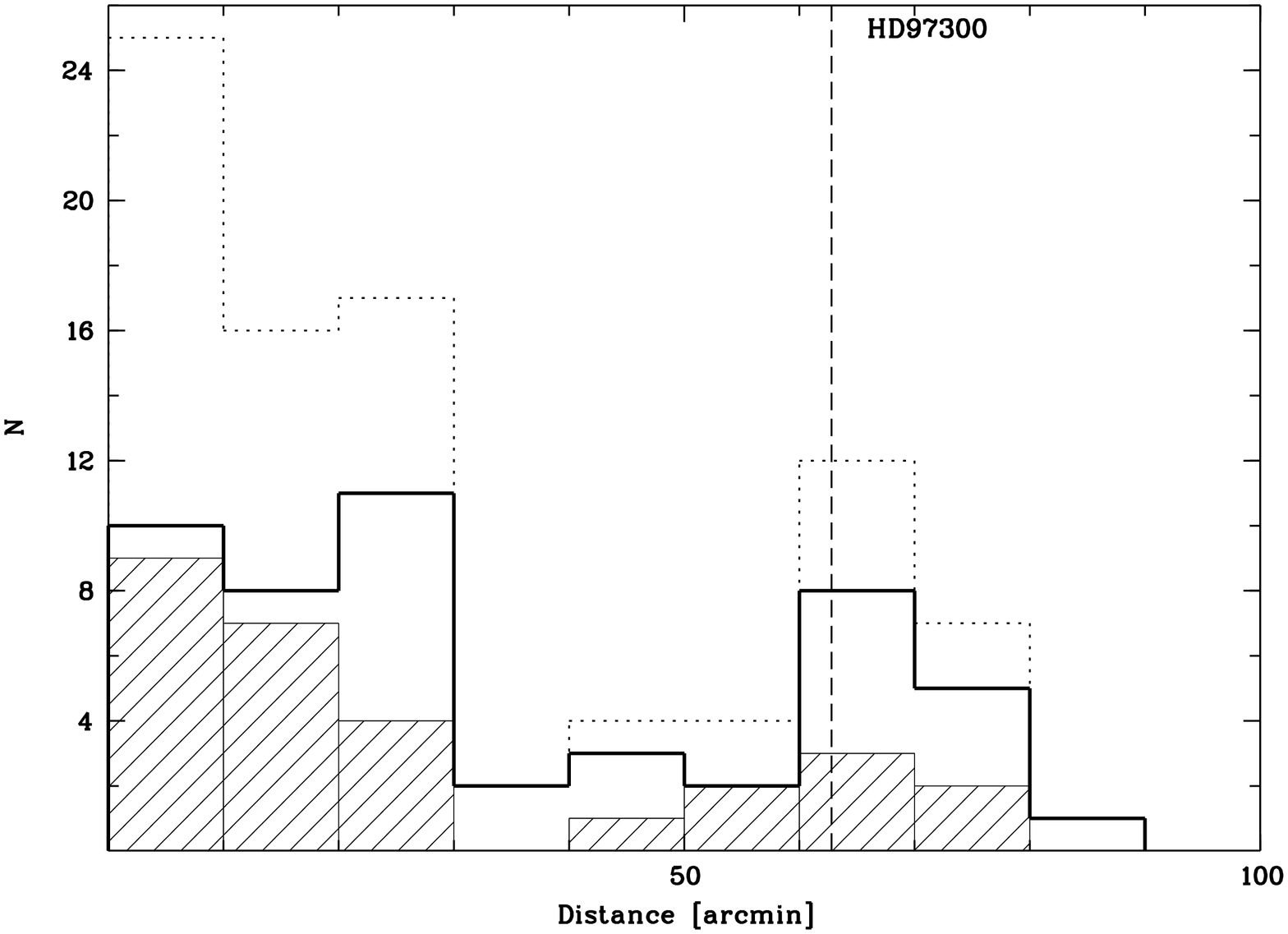} 
   \includegraphics[width=6cm, angle=-90, bb= 50 70 550 775]{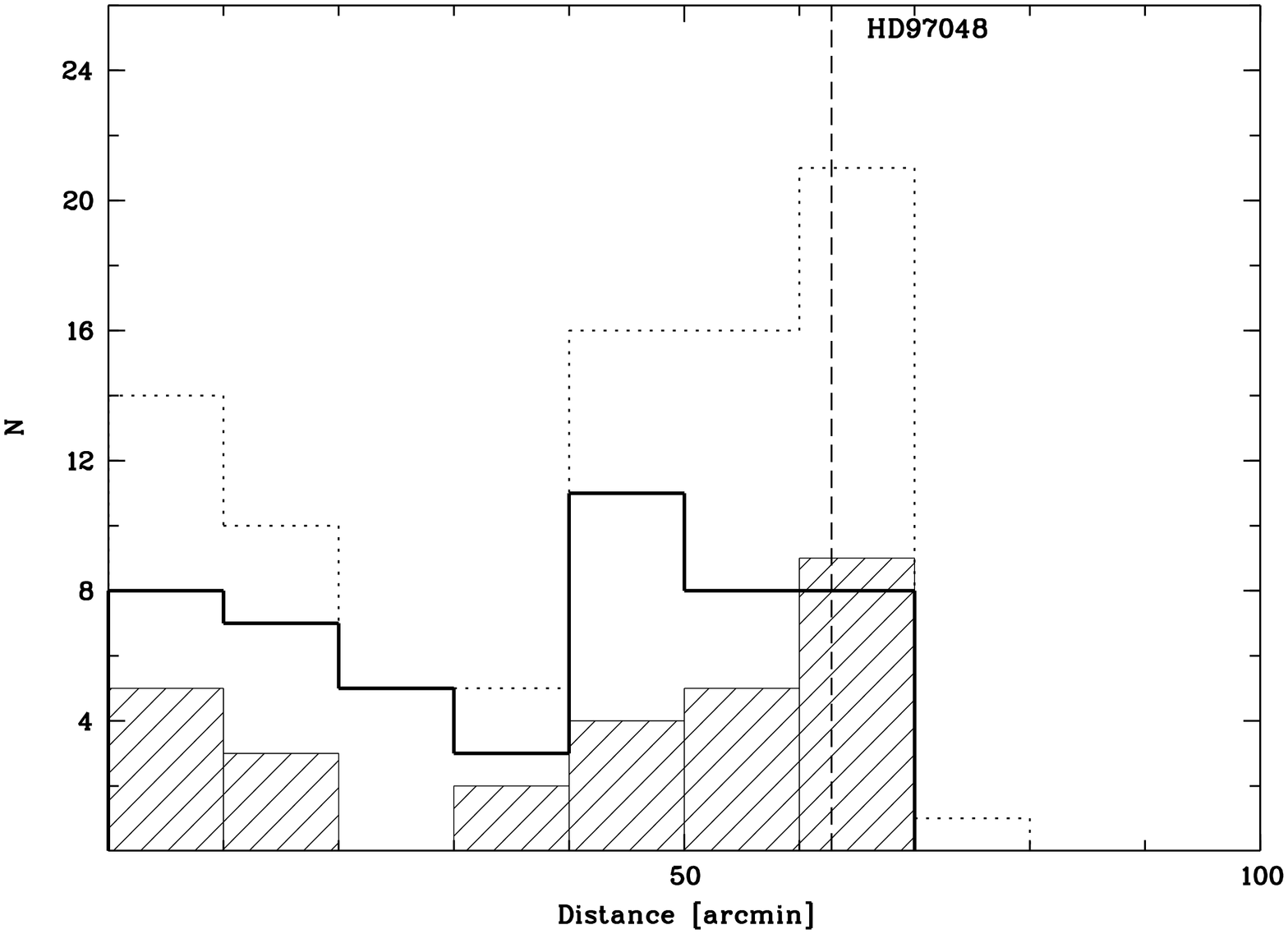}
  \caption{Distribution of the low-mass Cha~I members according to the 
distance
           to the intermediate mass stars HD~97048 (upper panel) and HD~97300
           (lower panel). The blank histograms show the distribution of the
           very low-mass stars; the hashed histograms, of the brown dwarfs and
           faint brown dwarf candidates. The general histograms for all these
           objects are showed in dotted lines. The distance to the other
           intermediate mass star is also indicated.}
              \label{fig:objdist}
    \end{figure}

%
\section{Spatial Distribution}\label{sec:dist}

        Fig.~\ref{fig:chaifield} shows the positions of our objects the 
Chamaeleon~I region, as well as those of the previously known brown dwarfs and 
very low-mass stars from \cite{comeron99} and \cite{comeron00}. The
distribution of the brown dwarfs and brown dwarf candidates resembles that of
the low-mass stars. All these objects are mostly located in the two cloud cores
containing the two intermediate mass stars HD~97048 and HD~97300 and at the 
boundaries of a third core, apparently the densest according to C$^{18}$O 
observations (Mizuno et al. \cite{mizuno99}). This core lies between the other 
two and is seen as a dark region devoid of stars in the optical images. Far 
from the three cloud cores, few objects are found. 

        To better quantify this observed distribution, we have counted the
number of objects at different distance bins from the stars HD~97048 and
HD~97300. Fig.~\ref{fig:objdist} shows the histograms derived from these
counts.\footnote{\footnotesize Although these counts might be somehow
incomplete, due to the fact that our studied area is larger in the North-South
than in the East-West direction, we do not expect a significant contribution
from objects beyond the survey boundaries because the cloud is mostly covered
by our observations (see e.g. Mizuno et al. \cite{mizuno99}). Thus, we do not
correct for boundary effects, which could introduce a new bias in the analysis
given that such corrections are generally based on the assumption of a random
distribution beyond the boundaries.}
We do see that the distributions of both stars and brown dwarfs peak near the 
positions of the intermediate mass stars. In general, the objects seem to be 
more clustered near HD~97048 (a Herbig Ae/Be star) than near HD~97300 (a likely
zero-age main sequence A star). A third peak is observed in both cases at the 
distance corresponding to the boundaries of the third cloud core between both 
stars (at around 30$\arcmin$ from HD~97048 and 40$\arcmin$ of HD~97300). 
Although some of the objects in the bins could actually be very far away from 
this core, the peak is significant because it appears when counting from both 
stars.

\section{Binary Objects}\label{sec:bin}

   \begin{figure}
   \centering
   \includegraphics[width=6cm, angle=0, bb= 50 70 550 775]{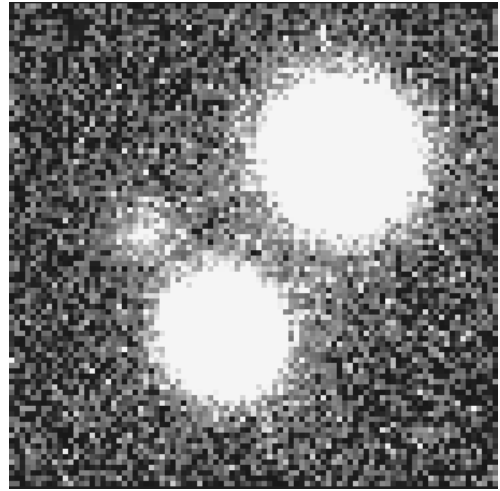} \hfill
     \caption{
                Detail of the long I exposure showing 
                the possible multiple 
                system formed by the star ChaI~405 (upper object), the brown 
                dwarf candidate ChaI~404 (lower  object) and a very faint 
                object towards the north-east of ChaI~404. The two
                brightest components lie at a separation of about 9$\farcs$5.
                This system is found at the position of the ISOCAM source
                ISO~250. 
                (North is up and east is left in this image.)
                }
         \label{fig:binary}
    \end{figure}

        To look for possible binary pairs among our objects, we adopt an upper
separation  cutoff of 12$\arcsec$ (about 1800 AU). This value was determined by
estimating the average surface density in our surveyed area. Then the distance
was computed at which a given object should have a single neighbour if there
were a uniform distribution on the whole field; this distance turned out to be
about 12$\farcs$2. Previous studies of visual binaries in Chamaeleon~I (e.g.
Reipurth \& Zinnecker \cite{reipurth93}; Brandner et al. \cite{brandner96})
also used an upper limit of 12$\arcsec$. Indeed, inspection of our images 
shows that the probability of finding a neighbouring object is notably 
increased at a slightly larger separation.

        Most of our objects are found in isolation, or at least do not have 
neighbours down to the distances resolvable from our images. In very few cases
one of our objects is found near a previously known cloud member: ChaI~625,
classified as a transition object (spectral type M6), is identified as the
brightest component of the visual binary Hn12 (Schwartz \cite{schwartz91}). The
M4 star ChaI~408 is located at about 14$\arcsec$ of the reported position for
the star \object{Sz~37} (which lies beyond our survey boundaries), and is
probably not physically bound to it.

        As mentioned in the previous section, the pair formed by ChaI~404 and
ChaI~405 lie whithin a separation of $\sim 9\farcs5$. If it were a true cloud
member, ChaI~404 would be a brown dwarf candidate with an estimated spectral 
type of M6.5. ChaI~405 has  spectral type M5.5 and is most probably a very 
low-mass star. This system could  still have a third member at about 
5$\arcsec$ ($\sim$750~AU) of ChaI~404. This object is only visible in our 
longest I exposure, with a measured magnitude of I$\simeq$19~mag. Considering 
our detection limits, it thus seems to have a very red colour, R--I$\gtrsim$3, 
which would fit the empirical isochrone for the Chamaeleon members shown in 
Fig.~\ref{fig:richai}. If this faint object indeed belonged to the cloud, it 
could be a very low-mass companion (a very low-mass brown dwarf or a giant 
planet) of this possible binary system. We show this possible multiple system 
in Fig.~\ref{fig:binary}.

        Another wide pair with a separation of $\sim$10$\arcsec$ is formed by 
ChaI~608 and ChaI~609. The latter is a star of spectral type M5 according to
our classification. Its location corresponds to the infrared-detected star
Baud~38  (Baud et al. \cite{baud84}). On the other hand, the position of the
ROSAT X-ray source CHXR~30 (Feigelson et  al. \cite{feigelson93}) is well
matched with that of ChaI~608. As mentioned in Sect.~\ref{sec:bdchai}, this 
object had a slightly too blue (R--I) colour to be included in our original 
candidate selection. We have found only one previous reference for a double 
object in a recent paper (Carpenter et al. \cite{carpenter02}), though already 
Feigelson et al. reported an offset of $7\arcsec$ between the positions of the
infrared and the X-ray sources. However, in our own inspection of the ROSAT 
data (L\'opez Mart\'{\i}, Stelzer \& Neuh\"auser \cite{lm02}), we do find a 
bright elongated source at the position of this binary candidate, meaning that 
the X-ray source could actually be double and include both ChaI~608 and 
ChaI~609.

        The pair formed by ChaI~735 and ChaI~737 (spectral types M5.5 and M8,
respectively), if actually bound, would  be a good candidate for a wide
star-brown dwarf system. However, these objects lie at a separation of
$\sim14\farcs5$, so they have a high probability of not being physically
connected. 

        We also checked our and the previously known objects for faint or very
close  companions that could have been missed by the automatic object search. 
We only find a possible companion near a brown dwarf candidate, ChaI~425
(spectral type M7.5), only visible in the longest I exposure as a faint 
''tail`` to this object. It was not possible to get the photometry of this 
possible companion from the subtracted image, due to the residuals left
by the subtraction of the brightest object (which is almost saturated in the
longest exposure). On the other hand, as explained in Sect.~\ref{sec:sptchai}, 
ChaI~456 is probably an unresolved stellar binary. ChaI~726 (a brown dwarf 
candidate of spectral type M8.5) seems to be slightly elongated in the 
north-south direction and could be a very close binary as 
well.\footnote{\footnotesize Recalling the discussion from 
Sect.~\ref{sec:sptchai}, if this were the case, the individual components of 
this binary system would probably have earlier spectral types than estimated 
from our study, and they might not be brown dwarfs.}

Although in several cases faint objects are seen near our identified cloud 
members, their (R--I) colours (when both I and R photometry are available)
seem to indicate that they belong to the background. Also some of the 
previously known brown dwarf candidates have faint neighbours visible in our 
images. These objects have recently been studied by Neuh\"auser et al. 
(\cite{neuhauser02}), who conclude from their infrared colours that they are 
unlikely to belong to the Chamaeleon~I cloud, and thus to be bound to the brown
dwarf candidates. To confirm if all these objects have close companions, 
spectroscopic observations as well as more imaging observations are needed.

        From this analysis we conclude that most of our objects are not members
of (wide) binary systems. Hence, binarity (down to the separations that we can
prove) should not significantly change the Chamaeleon~I mass function.

%
\section{Accretion Processes}\label{sec:accr}

   \begin{figure}
   \centering
  \includegraphics[width=6cm, angle=-90, bb= 50 70 550 775]{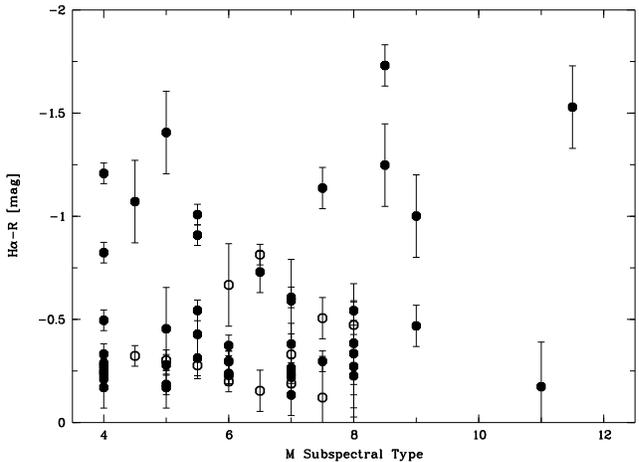} 
      \caption{Measured (H$\alpha$--R) colour index versus spectral type.
      Symbols as in Fig.~\ref{fig:richai}. The strongest emitters are among 
the
      objects with the latest spectral types.}
         \label{fig:haspt}
   \end{figure}
%

\subsection{H$\alpha$ Emission}\label{sec:hal}

        The detection of H$\alpha$ emission in young objects is commonly
regarded as indicative of an ongoing accretion process, and/or a measure of 
chromospheric activity. As explained in Sect.~\ref{sec:bdchai}, there is a 
relation between the H$\alpha$ equivalent width EW(H$\alpha$) and the 
(H$\alpha$--R) colour index for the objects studied by \cite{comeron00} 
(Fig.~\ref{fig:eqwi}). Since our new very low-mass stars and brown dwarfs
have similar (H$\alpha$--R) colours, they should also have similar H$\alpha$
equivalent widths, with values ranging between 10 and 30~\AA. In TTS, such
values are indicative of a significative accretion process.

        In Fig.~\ref{fig:eqwi} there is one remarkable outlayer: 
Cha\,H$\alpha$\,1. This bona-fide brown dwarf should have an equivalent width 
of $\sim 20$~\AA~ according to its (H$\alpha$--R) colour. However, it has a 
reported equivalent width between 34 and 99~\AA\, (\cite{comeron99}; 
Neuh\"auser \& Comer\'on \cite{neuhauser99}; \cite{comeron00} --the plotted 
data), being by far the strongest (and most variable) emitter of the whole 
data set. Natta \& Testi (\cite{natta01}) showed that Cha\,H$\alpha$\,1 is 
surrounded by a disk with similar characteristics to the ones around the more 
massive TTS. However, they also report a disk around Cha\,H$\alpha$\,2 and 
Cha\,H$\alpha$\,9, whose H$\alpha$ emission does not show a strong deviation 
from the general correlation in Fig.~\ref{fig:eqwi}. On the other hand, 
Cha\,H$\alpha$\,1  also has strong X-ray emission (Neuh\"auser \& Comer\'on 
\cite{neuhauser98}), probably an indicator of coronal magnetic activity. 
Therefore, its ``abnormal'' H$\alpha$ emission could also be related to 
magnetic (chromospheric) activity, or be the result of the combination of both 
accretion and chromospheric activity. A distinction between the two processes 
possibly leading to the strong H$\alpha$ emission, however, cannot be made 
with the available data. 

Recent observations show that some objects with photometric variability are 
characterized by a large H$\alpha$ equivalent width (Eisl\"offel \& Scholz 
\cite{eisloffel02}). However, no photometric variability of Cha\,H$\alpha$\,1 
has been reported so far, although observations from different epochs are 
available (Neuh\"auser \& Comer\'on \cite{neuhauser99}; \cite{comeron99}; 
\cite{comeron00}; Neuh\"auser et al. \cite{neuhauser02}). A study of the 
variability of some previously known brown dwarfs in Chamaeleon~I was recently 
performed by Joergens et al. (\cite{joergens03}), but these authors did not 
consider Cha\,H$\alpha$\,1 due to its poor S/N in their images. On the other 
hand, they find quite long photometric periods for ChaI\,H$\alpha$\,2 and 
ChaI\,H$\alpha$\,3 (3.2 and 2.2 days, respectively), while no variability was 
detected for ChaI\,H$\alpha$\,4, 5, 8 and 12. All these objects are included in
Fig.~\ref{fig:eqwi}. Since Cha\,H$\alpha$\,1 is the only one showing a 
different behaviour in its H$\alpha$ emission, it seems unlikely that it is 
connected to (long) photometric variability.

        Fig.~\ref{fig:haspt} shows the measured (H$\alpha$--R) colours for all
our objects with estimated spectral type. Although no clear systematic
behaviour is observed, interestingly, many of the objects with (H$\alpha$--R)
colours smaller than --1 (what roughly corresponds to equivalent widths greater
than 30~\AA) have spectral types later than M7. Thus, they are probably brown
dwarfs. Despite the larger photometric errors towards later spectral types,
i.e. fainter objects, it is remarkable that almost all the objects with
spectral type later than about M8 are found to have such extreme (H$\alpha$--R)
colours. However, the low number of objects does not allow us to state firmly
whether this is a typical property of substellar objects. Given that they are
just a small subset of our sample (14\%), it may be that these objects were
observed near a maximum of activity. It should be noted that this result may 
simply be a consequence of the sensitivity limit of our survey. Low luminosity 
objects showing moderate to low H$\alpha$ emission may appear below the 
completeness magnitude and close to the detection limit of the H$\alpha$ 
images, while on the contrary objects with strong emission are detected more 
easily.

Very strong H$\alpha$ emission has also been reported for a brown dwarf 
candidate in  \object{Corona Australis} (Fern\'andez \& Comer\'on 
\cite{fernandez01}). This object, \object{LS-RCrA~1} (M6.5), displays a very 
particular spectrum with spectral features of accretion and mass loss. This 
could also be the case of our strong H$\alpha$ emitters, as they probably have 
disks (see Sect.~\ref{sec:disks} below). Such a result would be surprising 
because, if brown dwarfs form from the collapse of a molecular cloud core, they
are expected to have very low accretion rates. Otherwise, they would eventually
accrete enough mass to start the hydrogen burning process. This is in 
contradiction with the observation of strong H$\alpha$ emission. We note, 
however, that this emission can seem  stronger in late-type objects simply due 
to a lower continuum. Moreover, indications of accretion have not been found in
all substellar objects showing a prominent H$\alpha$ emission line. This is the
case e.g. for \object{S~Ori~71}, a substellar candidate member of the 
$\sigma$~Orionis cluster (Barrado y Navascu\'es et al. \cite{barrado02}). These
authors discuss other possible causes of the H$\alpha$ emission, such as 
chromospheric activity or mass exchange between the components of a binary 
system. Complementary spectroscopic observations of our objects will certainly 
help to clarify the origin and characteristics of their H$\alpha$ emission.

   \begin{figure}
   \centering
  \includegraphics[width=6cm, angle=-90, bb= 50 70 550 775]{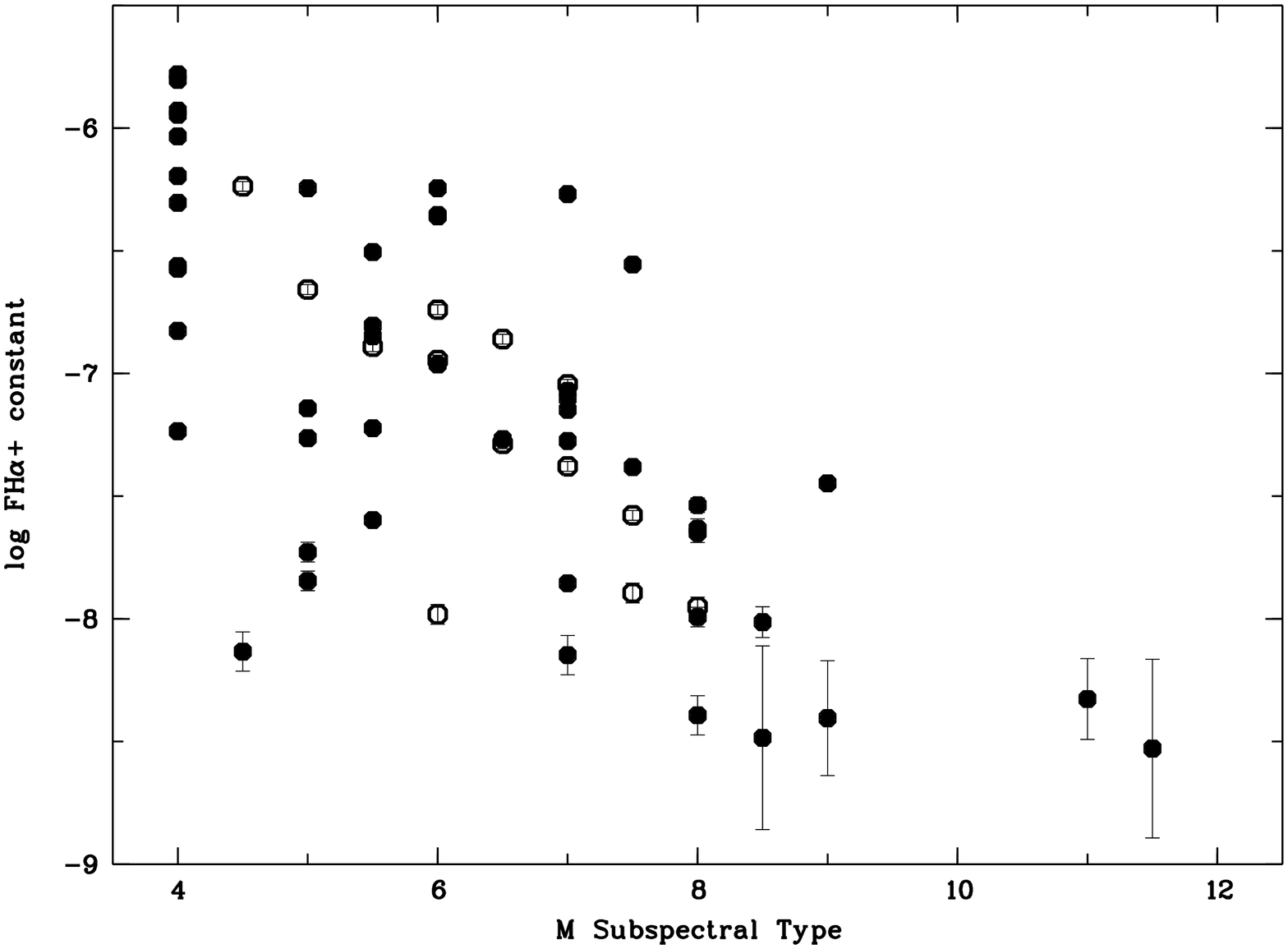}\hfill 
  \includegraphics[width=6cm, angle=-90, bb= 50 70 550 775]{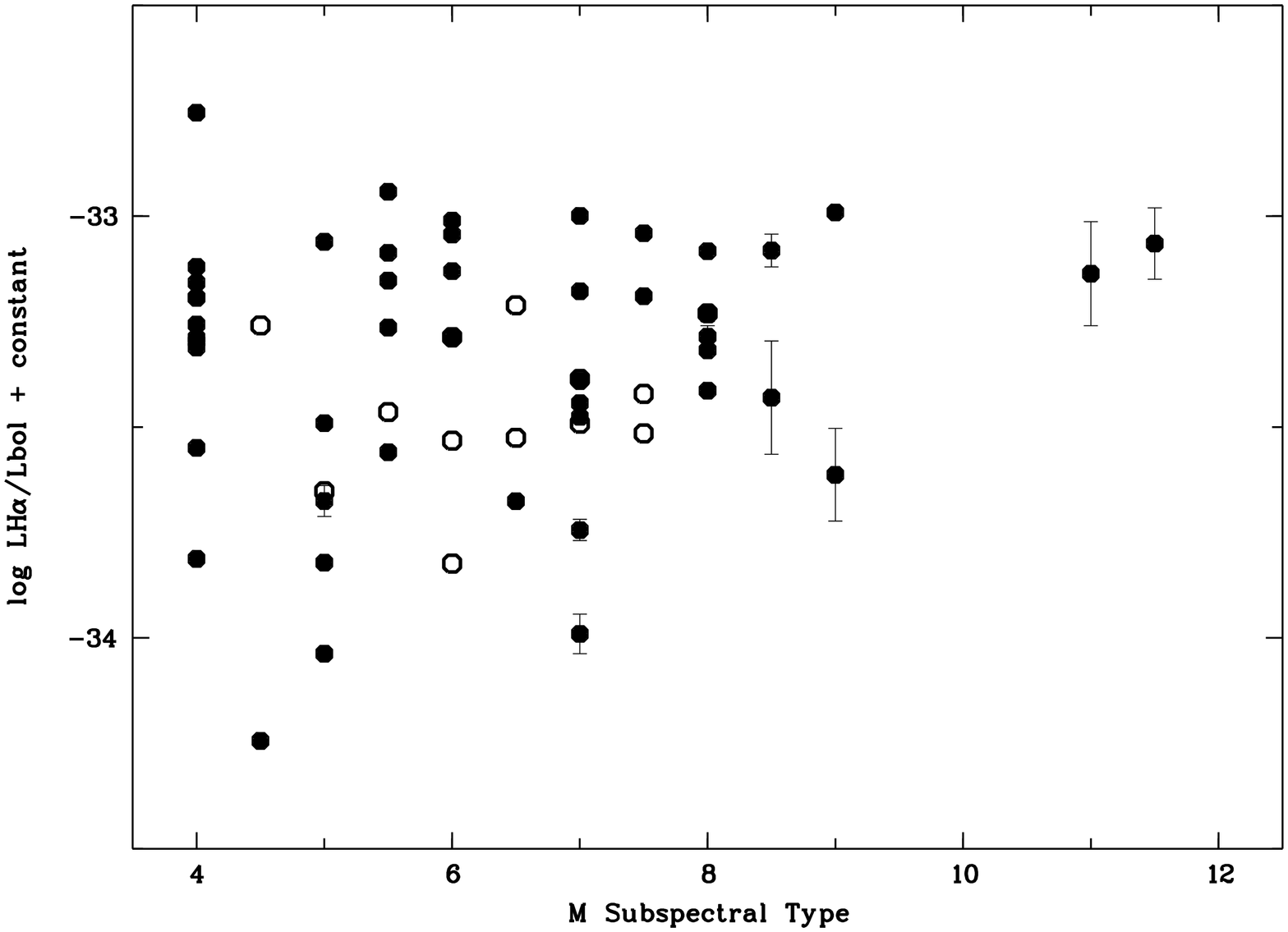}\hfill 
      \caption{\footnotesize
                \emph{Upper panel:}
                 Measured H$\alpha$ pseudoflux versus spectral type for 
                our objects in Chamaeleon~I. Symbols as in 
                Fig.~\ref{fig:richai}. The flux clearly decreases with later 
                spectral type.
                \emph{Lower panel:} $\log$L$_{\mathrm{H}\alpha}$/L$_{bol}$ 
                versus spectral type for our objects in Chamaeleon~I.
               }
         \label{fig:fhaspt_chai}
   \end{figure}

        We also analysed the behaviour of the H$\alpha$ flux with the spectral
type. Since no standards were available for the H$\alpha$ filter, we computed 
a ``pseudoflux'' using the relation: 

\begin{center}
\begin{equation}
\label{eq:fha}
\tilde{\mathrm{F}}_{\mathrm{H}\alpha}=
\mathrm{F}_{\mathrm{H}\alpha}/\mathrm{F}_0=
10^{-m_{\mathrm{H}\alpha}/2.5}, 
\end{equation}
\end{center}

\noindent
where $m_{\mathrm{H}\alpha}$ is the measured magnitude in the H$\alpha$ filter.
The difference between the logarithm of the pseudoflux \~F$_{\mathrm{H}\alpha}$
and that of a hypothetical absolute flux F$_{\mathrm{H}\alpha}$ is then a
constant value, log~F$_0$, depending on the reference photometric system.  

The upper panel in Fig.~\ref{fig:fhaspt_chai} shows 
log~\~F$_{\mathrm{H}\alpha}$ versus M subspectral type for Chamaeleon~I. A 
roughly linear decrease of the flux with later spectral type is clearly seen, 
consistent with the progressive faintness of the objects. There is no 
indication for a change at the stellar/substellar boundary. The apparent 
saturation for the latest spectral types is most probably due to the 
sensitivity limit of our H$\alpha$ survey. 

        A decrease of the chromospheric activity with the spectral type has
been reported by Gizis et al. (\cite{gizis00}) for older M- and L-type dwarfs.
Also Zapatero Osorio et al. (\cite{zo02a}) found a similar behaviour in a 
sample of very low-mass members in the young $\sigma$~Orionis cluster.
We note, however, that these authors used 
L$_{\mathrm{H}\alpha}$/L$_\mathrm{bol}$ instead of the H$\alpha$ flux in their 
study, and thus our result is not directly comparable with theirs. 
L$_{\mathrm{H}\alpha}$/L$_\mathrm{bol}$ is probably a better activity 
indicator, since it does not depend on the distance or the radius of the 
objects. 

The lower panel in Fig.~\ref{fig:fhaspt_chai} is a plot of
$\log$\~L$_{\mathrm{H}\alpha}$/L$_\mathrm{bol}$ versus the spectral type for 
our young objects in Chamaeleon~I. To obtain the bolometric luminosities, we 
have proceeded as in \cite{comeron00}, using (R--I) intrinsic colours from 
Kenyon \& Hartmann (\cite{kenyon95}) and Zapatero Osorio et al. (\cite{zo97}). 
Interestingly, we do not see a decrease in the H$\alpha$  emission in these 
data. The situation reminds us of the results of Mokler \& Stelzer 
(\cite{mokler02}), who studied the activity of a large sample of young very 
low-mass objects and found no decrease in L$_\mathrm{x}$/L$_\mathrm{bol}$ down 
to spectral type M7, although a decrease in L$_\mathrm{x}$ was clearly seen in 
their data.

Zapatero Osorio et al. (\cite{zo02a}) noted that young objects displayed, on 
average, higher H$\alpha$ emission than their older field counterparts, the 
difference being higher for L-type objects. They explained this result as a 
consequence of the decline of magnetic activity and/or mass accretion with age.
Such a direct comparison between the H$\alpha$ luminosities is not possible 
with our data, however, because of the lack of an absolute H$\alpha$ 
photometric system.

   \begin{figure}
   \centering
   \includegraphics[width=6cm, angle=-90, bb= 50 70 550 775]{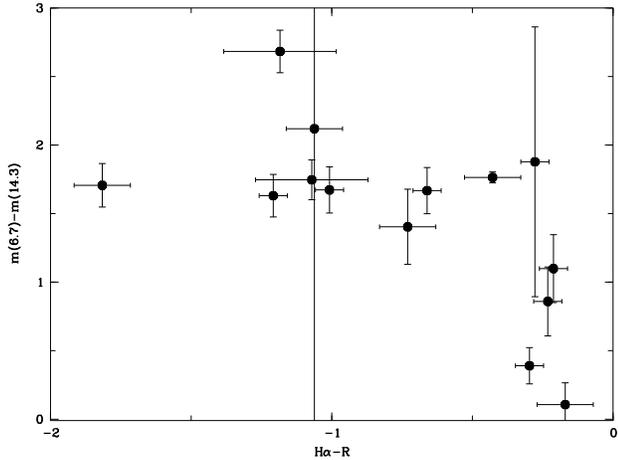}
      \caption{(H$\alpha$--R, m(6.7)--m(14.3)) colour-colour diagram for the
                objects in our sample detected by ISOCAM in both passbands. The
                objects with very strong H$\alpha$ emission have 
                mid-infrared excess.}
         \label{fig:isoha}
   \end{figure}
%

\subsection{Evidence for Disks}\label{sec:disks}

        Fig.~\ref{fig:isoha} is a plot of the colour index  defined by the two
ISOCAM passbands, m(6.7)--m(14.3) (computed with Eqs.~\ref{eq:isomag1} and 
\ref{eq:isomag2}), versus our measured (H$\alpha$--R) colour index. All our 
objects with strong H$\alpha$ emission from their (H$\alpha$--R) colours are 
found to have mid-infrared excess according to Persi et al. (\cite{persi00}). 
The detection of such an excess from young stellar objects is a well-known 
observational signature of a circumstellar disk. Hence, H$\alpha$ emission is 
related to the presence of an accretion disk around these objects.

        A similar proportion of the stars and brown dwarfs in our sample
(around 30\% and 37\%, respectively) have a reported mid-infrared excess. On
the other hand, while around 52\% of the stars and 67\% of the transition
objects show near-infrared excess, only about 37\% of the brown dwarfs do.
Interestingly, mid- and near-infrared excess do not seem to be correlated: of
the brown dwarfs with mid-infrared excess, just around 10\% show also
near-infrared excess. This percentage slightly increases for transition 
objects and low-mass stars (17\% and 21\%, respectively). However, the total 
proportion of objects detected with both kinds of excesses is only around 
13\%. This is in good agreement with the results of \cite{comeron00}, who found
that several of the previously known brown dwarfs and brown dwarf candidates in
Chamaeleon~I (selected from an H$\alpha$ survey) showed mid-infrared excess, in
spite of lacking near-infrared excesses. 

         For TTS, the lack of near-infrared excess is usually attributed to an
inner hole in the circumstellar disk. However, strong H$\alpha$ emission is
normally interpreted as a signature of the accretion process onto the surface 
of the object, what requires the presence of an inner disk. In a recent 
paper, Natta \& Testi (\cite{natta01}) show that the observed spectral energy 
distribution of three of the previously known brown dwarfs and brown dwarf
candidates in Chamaeleon~I can be explained if they are surrounded by accretion
disks. They remark that, in the case of brown dwarfs, the presence of disks is
difficult to infer from near-infrared photometry alone, where the emission is
largely dominated by the photosphere. Also Oliveira et al. (\cite{oliveira02}),
who performed infrared observations of a sample of very low-mass objects in the
$\sigma$Orionis cluster, conclude that only a very small proportion of them 
show significant near-infrared excess. Despite this negative result, some of 
the low-mass objects in this cluster show H$\alpha$ emission and other spectral
features characteristic of an accretion process (Zapatero-Osorio et al. 
\cite{zo02}). Indeed, a recent $L$-band study has increased the disk frequency 
of the low-mass $\sigma$Orionis members up to about 50\%, remarkably larger 
than inferred from the $K$-band excess alone (Oliveira et al. 
\cite{oliveira03}).

        In any case, the indications for the presence of disks around many of
our brown dwarfs and brown dwarf candidates favours the hypothesis that they
formed in a similar way as TTS, and leave open the possibility of brown dwarfs
forming planetary systems.

        Although mid-infrared photometry is available for two of our faintest 
objects (ChaI~607 and ChaI~441), neither of them shows an excess according to 
the ISOCAM observations. However, ChaI~607, and also ChaI~428, have a 
near-infrared excess. As argued in Sects.~\ref{sec:sptchai} and \ref{sec:ir}, 
both ChaI~607 and ChaI~428 may be very reddened stars. Their observed excess 
would then be due to their high extinction (with values in the $J$ band of 4.9 
and 3.9, respectively, according to G\'omez \& Mardones \cite{gomez03}). It 
seems thus probable that also ChaI~441 is an extincted star, rather than a 
low-mass brown dwarf. 

   \begin{figure}
   \centering
   \includegraphics[width=6cm, angle=-90, bb= 50 70 550 775]{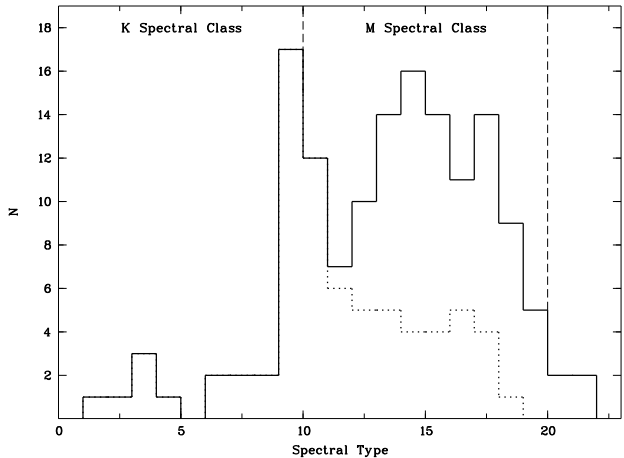}
      \caption{Histogram showing the distribution of spectral types for the
      identified low-mass stars and brown dwarfs in Chamaeleon~I.  
      The dotted line is the distribution for the objects studied by previous
      authors (Gauvin \& Strom \cite{gauvin92}; Lawson et al. \cite{lawson96};
      \cite{comeron99} and \cite{comeron00}).
      The local minimum at M2-M3
      may be due to the incompleteness of our sample in this range (see text).
      }
         \label{fig:sptdist}
   \end{figure}
%
%
\section{The Mass Function in Chamaeleon~I}\label{sec:chaimf}

        The task of constructing the substellar initial mass function is not 
an easy one because the theoretical models still present large uncertainties 
for such very young objects ($<10$~Myr). Especially relevant is the choice of 
initial conditions: The picture of non-accreting objects contracting from large
initial radii, as used in most of the models, is an idealized description, 
which can drastically affect the fundamental properties (luminosity and 
effective temperature) of objects younger than about 10~Myr (Baraffe et al. 
\cite{baraffe02}). From the observational point of view, the main problems are 
the uncertainties in the luminosities and spectral types for these young, often
highly extincted objects. Moreover, recent observations show that very young 
brown dwarfs can be variable (e.g. Eisl\"offel \& Scholz \cite{eisloffel02}). 
Therefore, any attempt to infer an age or a mass from observable quantities for
these objects must be considered with caution.

	On the other hand, at these very early stages of evolution there is a 
good correlation between the spectral type (i.e. the effective temperature) 
and the mass. Thus, as a first step, we can study the spectral type 
distribution  of the Chamaeleon~I members shown in Fig.~\ref{fig:sptdist}. The
histogram includes all the new members with a determined spectral type from
Table~\ref{tab:chaiphot} (except the few dubious objects discussed in the 
previous sections), as well as the objects in the surveyed area with identified
spectral type from Gauvin \& Strom (\cite{gauvin92}), Lawson et al.
(\cite{lawson96}), \cite{comeron99} and \cite{comeron00}. The dotted line 
indicates the histogram for these previously known stars. Only the objects
with spectral types between K0 and L2 are shown. A numeric code is used: The
numbers 2 to 9 correspond to K0 to K7 objects, respectively; 10-19 indicate
spectral types from M0 to M9; and numbers later than 20 are asigned to the
L-type objects.  Few objects of spectral types earlier than K are known in the
cloud: Gauvin \& Strom report four objects of spectral type G, as well as the
intermediate-mass stars HD~97048 (B9) and HD~97300 (A0). 

	Two peaks are seen in the spectral type distribution. The first one 
corresponds to late K objects. Nonetheles, the decrease in the number of 
objects for the spectral types M0-M3 might not be a real characteristic of the 
region, but a consequence of the incompleteness of our sample in this range. We
recall that the (M855--M915) colour index saturates for spectral types earlier 
than M4; hence, twelve objects in our survey that have a spectral type earlier 
than M4 but lack a previous spectroscopic classification are not included in 
Fig.~\ref{fig:sptdist}. The second  peak is seen for spectral types around M5, 
approximately at the star/brown dwarf transition, indicating a decrease of the 
number of objects in the substellar regime. Taking M5 as the stellar/substellar
boundary, we count around 100 stars (including the ones from our study whose 
spectral type is unknown) and about 51 brown dwarfs and brown dwarf candidates.
Hence, the number of stars in the surveyed area is roughly doubling that of 
brown dwarfs. It must be noted, however, that this is just a lower limit to the
actual number of brown dwarfs in the cloud, because some low-mass objects 
without (detectable) H$\alpha$ emission  will be missed due to our selection 
criteria. Moreover, extincted brown dwarfs are more difficult to detect than 
stars, especially at optical wavelengths.

Taking into account the uncertainties in our absolute photometry and spectral
types (recall discussions in Sect.~\ref{sec:cal} and \ref{sec:sptchai}), 
especially for objects earlier than M4, we can now construct the mass function
in Chamaeleon and give an estimation for the value of the exponent $\alpha$ of
the usual approximation for the mass function:

\begin{equation}\label{eq:imflin}
d\mathrm{N}/d\mathrm{M} \sim \mathrm{M}^{-\alpha} 
\end{equation}

For this calculation, we considered all known members between spectral types K7
and M7, for which our survey should be complete. According to the models of 
Baraffe et al. (\cite{baraffe98}), this range should correspond to masses of 
1.2 to 0.025\,M$_{\sun}$ at the age of 1~Myr. We splitted the objects into two 
bins of stars (spectral types K7--M4, with masses of about 
1.2--0.075\,M$_{\sun}$) and substellar objects (spectral types M5--M7, with 
masses of about 0.075--0.025\,M$_{\sun}$). We did not take smaller bins because
of the uncertainty in the classification for some of our objects with spectral 
types earlier than M4. By selecting two bins, we can include all the identified
Chamaeleon~I members in the calculation, even though we do not know their 
spectral types very accurately. There are 88 objects in the first and 39 in 
the second bin.\footnote{\footnotesize We recall that the twelve objects with 
uncertain spectral type are not plotted in Fig.~\ref{fig:sptdist}.} 
With these data, we derive an exponent of the mass function of

\begin{equation}\label{eq:alpha}
\alpha = 0.6 \pm 0.1
\end{equation} 

As mentioned above, some cloud members may have been missed because they  do
not have  (measurable) H$\alpha$ emission. To account for this possibility, we
repeated  the calculation several times including all the candidates from our R
and I  photometry that would belong to the first, second and both bins if they
were  indeed cloud members. These considerations lead us to the quoted error
of  0.1 for  the index of the mass function.

Our obtained index agrees surprisingly well with the value obtained by Moraux
et al. (\cite{moraux03}) for the Pleiades cluster ($0.60\pm0.11$). As these
authors point out, a similar value has been derived also for other clusters of 
similar age, thus older than Chamaeleon~I. These results are also consistent 
with the values of $0.8 \pm 0.4$ and $0.7 \pm 0.2$ which B\'ejar et al. 
(\cite{bejar01}) and Tej et al. (\cite{tej02}) found for the similarly old 
clusters \object{$\sigma$\,Orionis} and IC\,348, respectively. Hence, the shape
of the IMF at the substellar boundary seems to be similar in all these young 
regions, regardless of their age and crowdedness.
 
A previous estimate of the index of the  substellar IMF in Chamaeleon~I had
been done by \cite{comeron00}. They obtained a value $\alpha = 1.1$, larger
than ours, implying a steeper increase of the mass function towards lower
masses. The discrepancy might be due to the low number of objects in their
study. We recall that the area covered by \cite{comeron00} is just 15\% of our
whole survey, and that they studied a region particularly dense in very
low-mass objects.

\section{Constraints on Formation Models}\label{sec:bdfor}

        The spatial distribution of the brown dwarfs in Chamaeleon~I is found
to be a natural continuation of the T~Tauri stars towards lower masses 
(see Sect.~\ref{sec:dist}). Stellar and substellar objects also appear to have 
similar H$\alpha$ emission properties (Sect.~\ref{sec:hal}). These results, 
together with the presence of accretion disks around some brown dwarfs 
(Sect.~\ref{sec:disks}) and the low number of possible star-brown dwarf systems
(Sect.~\ref{sec:bin}), lead us to assume that they most probably formed in the 
same way as TTS (that is, from the gravitational collapse of a molecular cloud)
rather than in a circumstellar disk like planets do. 

Moreover, the fact that the low-mass objects tend to be concentrated in 
particular areas seems to indicate that they formed very near the places where 
we currently see them. We note that a star with a tangential velocity of 
1~km\,s$^{-1}$ would have moved 3~pc in 3~Myr, which corresponds to about 
1$\fdg$1 at the distance of Chamaeleon~I. Thus, if most of the brown dwarfs
had been stellar embryos ejected from the stellar systems where they formed
with an initial velocity of 1~km\,s$^{-1}$, at the age of Chamaeleon~I we
should  see them randomly distributed over the whole surveyed region, which is
not the case. The typical velocity of our observed young stars and brown 
dwarfs cannot be higher than 0.2-0.4~km\,s$^{-1}$ at most. This value is also 
remarkably lower than the average ejection velocities of 2~km\,s$^{-1}$ 
predicted by the simulations of Bate et al. (\cite{bate03}). Joergens \& 
Guenther (\cite{joergens01}) report a \emph{radial} velocity dispersion of 
2~km\,s$^{-1}$ for several previously known brown dwarf and brown dwarf 
candidates around HD~97048. They argue, however, that this value is 
significantly smaller than the radial velocity dispersion of the TTS in 
Chamaeleon~I (3.6~km\,s$^{-1}$). On the basis of these considerations, it seems
very unlikely that the brown dwarfs in Chamaeleon~I have been ejected from 
multiple systems.

        From the observed spatial distribution one may think that the existence
of low-mass objects in Chamaeleon~I could be somehow connected with the 
presence of the two intermediate mass stars. Dissipation by stellar winds from 
a nearby hot massive star could prevent the further growth of a stellar core, 
which would then remain a failed star --a brown dwarf. This mechanism has been 
invoked recently to explain the high concentration of substellar objects in 
Orion (Lucas \& Roche \cite{lucas00}; Lucas et al. \cite{lucas01}). The stars 
in Chamaeleon~I, however, are not so hot and massive as the ones in Orion. 

        HD~97048, the one exhibiting a higher concentration of very low-mass 
neighbours, is a Herbig Ae/Be star, an object characterized by its strong
winds. But it is doubtful whether winds from Herbig Ae/Be stars are powerful
enough to evaporate the envelopes of surrounding star forming cores. This seems
even more unlikely to be currently happening in the case of HD~97300, which
might have reached the zero-age main sequence (e.g. Jones et al. 
\cite{jones85}). Indeed, according to Gauvin \& Strom (\cite{gauvin92}),
observations suggest that this part of the cloud is in a more evolved stage
than the one containing the other intermediate mass star. This is also
supported by the fact that the  cloud density, as inferred from CO
measurements, is significantly lower in  this core than in the other two
(Mizuno et al. \cite{mizuno99}). 

        Furthermore, another remarkable concentration of low-mass objects is 
seen at the boundaries of the third cloud core, which does not contain  any
intermediate mass star. This core has not been as well studied as the other
two. It is possible that we just observe the objects located in the outer part
of this core, and that objects deep inside it are too obscured to be detected
by our survey. Deeper observations are needed to clarify this point. If
embedded brown dwarfs were detected, it would certainly mean that environmental
conditions other than the presence of more massive stars are responsible for
the formation of the substellar objects in Chamaeleon~I. 

Finally, our results from Sect.~\ref{sec:chaimf} indicate that the mass 
function of Chamaeleon~I is rising towards lower masses. Hence, no matter how 
the brown dwarfs actually formed, this mechanism seems to be very efficient in 
producing substellar objects. The similar index $\alpha$ obtained for the mass 
functions of several young clusters and star forming regions hints to a similar
formation process in all these regions.

%
\section{Conclusions}\label{sec:concl}

        We performed a deep WFI survey of about 1.2~$\sq^{\circ}$ of the
Chamaeleon~I cloud to identify new very low-mass members of this star-forming
region. Candidates were selected from R and I photometry and their youth was
tested by means of their H$\alpha$--R colours. We showed that this colour is a
useful estimator of the H$\alpha$ emission rate.

        We developed a method to classify mid- to late M and early L-type
objects from photometric observations in the two intermediate-band filters 
M855 and M915. The colour defined by these two passbands, (M855--M915), is
correlated with the depth of the TiO and VO spectral features in these 
objects. In this way, we could provide the first spectral classification for
about 80\% of the objects in our sample.

        From our results, we conclude that the Chamaeleon~I dark cloud 
contains a substantial amount of very low-mass objects, with its spectral type
distribution showing a peak at  about the star/brown dwarf transition boundary.
Both kinds of objects have a similar spatial distribution, being placed
predominantly near the cloud cores. Such a distribution does not seem to be a
consequence of the presence of the two intermediate-mass stars, HD~97048 and
HD~97300, because a large number of cloud members are probably associated with
a dense core devoid of optical objects. We derive an index 
$\alpha = 0.6 \pm 0.1$ of the mass function in the range from 1.2 to 
0.025~M$_{\sun}$, very similar to the values obtained by other authors in other
young clusters and star forming regions.

        Both the observed spatial distribution and the scarce number of visual 
pairs among our objects seem to indicate that the brown dwarfs in Chamaeleon~I 
formed in a way similar to low-mass stars, and that they were not ejected from 
their parental systems, at least with high escape velocities (greater than
0.2-0.4~km\,s$^{-1}$). Moreover, the detection of H$\alpha$ emission connected 
to mid-infrared excess shows that at least some brown dwarfs probably have 
accretion disks, providing a further argument in favour of a star-like 
formation. On the other hand, near-infrared excess does not appear as a good
diagnostic tool to the presence of disks around very low-mass objects. 

The H$\alpha$ emitting properties of the stars and brown dwarfs are very 
similar. There is no evidence for a change across the substellar transition. 
Contrary to other authors, we do not see a decrease in 
$\log$L$_{\mathrm{H}\alpha}$/L$_{\mathrm{bol}}$ for spectral types M4-M9, 
meaning that the latest-type objects in this range could have at least as 
strong H$\alpha$ emission as earlier-type objects. All this hints to a common 
origin for the emission in both stars and brown dwarfs, probably accretion 
processes, since both kinds of objects have disks. If it is confirmed, this 
result would contradict the commonly accepted picture that objects with lower 
mass should have lower accretion rates (hence, weaker H$\alpha$ emission) than 
those with higher masses. However, with the available data we cannot exclude 
that other mechanisms, such as chromospheric activity, are (at least partially)
responsible for this emission. In this context, it is interesting that a 
similar result has been obtained by Mokler \& Stelzer (\cite{mokler02}) for the
X-ray emission of substellar objects.

%
\begin{acknowledgements} 

        We are very grateful to H.~Jahreiss for providing us with a list of
objects from the Gliese-Jahreiss catalogue to be used as spectral standards; 
and to I.~Baraffe, who calculated the medium-band filter isochrones for us. We 
kindly thank C.~A.~L.~Bailer-Jones for his help with the field selection, 
P.~Persi for the complete ISOCAM list, and K.~Tachihara for the information 
about the CO data. We also acknowledge the useful suggestions made by 
R.~Neuh\"auser and by an anonymous referee, as well as many valuable comments 
from F.~Comer\'on, A.~Fern\'andez Soto, J.~Fabregat, J.~M.~Torrelles and 
J.~M.~Marcaide. J.~E. thanks E.~Pompei and the 2.2-m team for their help with 
the observations and the taxi services for their walking-impaired visiting 
astronomer.

      We made use of the SIMBAD database, operated at the  
\emph{Centre de Donn\'ees astronomiques de Strasbourg (CDS)} in Strasbourg
(France). This work was supported by the German \emph{Deut\-sche
For\-schungs\-ge\-mein\-schaft, (DFG)}, projects numbers EI~409/7-1 and
EI~409/7-2. B.~L.~M. also thanks ESO for financial support through its Director
General's Discretionary Funds during her stay at the ESO Headquarters in 
Garching.

\end{acknowledgements}

%


\begin{thebibliography}{}

\bibitem[2002]{alcala02} 
Alcal\'a, J.~M., Radovich, M., Silvotti, R. et al. 2002, SPIE 4836, 406A

\bibitem[1989]{appenzeller89} 
Appenzeller, I. \& Mundt, R.\ 1989, \aapr \ 1, 291

\bibitem[2002]{armitage02} 
Armitage, P.~J.~\& Bonnell, I.~A.\ 2002, \mnras \ 330, L11 

\bibitem[1998]{baade98} 
Baade, D., Meisenheimer, K., Iwert, O. et al.\ 1998, 
\emph{The ESO Messenger} 93, 13

\bibitem[1998]{baraffe98} 
Baraffe, I., Chabrier, G., Allard, F., \& Hauschildt, P.~H.\ 1998, 
\aap \ 337, 403 

\bibitem[2002]{baraffe02} 
Baraffe, I., Chabrier, G., Allard, F., \& Hauschildt, P.~H.\ 2002, 
\aap \ 382, 563 

\bibitem[2001]{bn01} 
Barrado y Navascu\'es, D., Stauffer, J.~R., Brice\~no, C., et al. 2001, 
\apjs \ 134, 103    

\bibitem[2002]{barrado02} 
Barrado y Navascu{\' e}s, D., Zapatero Osorio, M.~R., Mart{\'{\i}}n, E.~L. 
et al. \ 2002, \aap \ 393, L85 

\bibitem[2003]{bate03} 
Bate, M.~R., Bonnell, I.~A. \& Bromm, V. 2003, \mnras \ 339, 577 

\bibitem[1984]{baud84} 
Baud, B., Beintema, D.~A., Wesselius, P.~R.,~et al.\ 1984, \apjl \ 278, L53 

\bibitem[1999]{bejar99} 
B{\' e}jar, V.~J.~S., Zapatero Osorio, M.~R., \& Rebolo, R.\ 1999, 
\apj \ 521, 671 

\bibitem[2001]{bejar01} 
B{\' e}jar, V.~J.~S., Mart\'{\i}n, E.~L., Zapatero Osorio, M.~R.,~et al.\ 
2001, 
\apj \ 556, 830 

\bibitem[1996]{bertin96} 
Bertin, E.~\& Arnouts, S.\ 1996, \aaps \ 117, 393 

\bibitem[1998]{bouvier98} 
Bouvier, J., Stauffer, J.~R., Mart\'{\i}n, E.~L., et al. 1998, \aap \ 336, 490 

\bibitem[1996]{brandner96} 
Brandner, W., Alcal\'a, J.~M., Kunkel, M. et al.\ 1996, \aap \ 307, 121 

\bibitem[1998]{briceno98} 
Brice{\~ n}o, C.~;., Hartmann, L., Stauffer, J., \& Mart{\' i}n, E.\ 1998, 
\aj \ 115, 2074 

\bibitem[1998]{cambresy98} 
Cambr\'esy, L., Copet, E., Epchtein, N., et al. \ 1998, \aap \ 338, 977 

\bibitem[2002]{carpenter02} 
Carpenter, J.~M., Hillenbrand, L.~A., Skrutskie, M.~F., \& Meyer, M.~R.\ 2002, 
\aj \ 124, 1001

\bibitem[2000]{chabrier00} 
Chabrier, G., Baraffe, I., Allard, F., \& Hauschildt, P.\ 2000, 
\apj \ 542, 464 

\bibitem[CRN99]{comeron99} 
Comer{\' o}n, F., Rieke, G.~H., \& Neuh{\" a}user, R.\ 1999, \aap \ 343, 477
(CRN99)

\bibitem[CNK00]{comeron00} 
Comer{\' o}n, F., Neuh{\" a}user, R., \& Kaas, A.~A.\ 2000, \aap \ 359, 269
(CNK00) 

\bibitem[1999]{delfosse99} 
Delfosse, X., Tinney, C.~G., Forveille, T.,~et al. \ 1999, \aaps \ 135 41 

\bibitem[2002]{eisloffel02} 
Eisl\"offel, J. \& Scholz, A. 2002, \emph{AG Abstract Series} 19, A15

\bibitem[2003]{eisloffel03} 
Eisl\"offel, J., Scholz, A., Mundt, R. et al. 2003, in preparation

\bibitem[1997]{epchtein97} 
Epchtein, N., de Batz, B., Capoani, L.~et al.\ 1997, 
\emph{The ESO Messenger} 87, 27 

\bibitem[1993]{feigelson93} 
Feigelson, E.~D., Casanova, S., Montmerle, T., \& Guibert, J.\ 1993, 
\apj \ 416 623 

\bibitem[1989]{feigelson89} 
Feigelson, E.~D. \& Kriss, G.~A. 1989, \apj \ 338, 262

\bibitem[2001]{fernandez01} 
Fern\'andez, M. \& Comer\'on, F. 2001, \aap \ 380, 264

\bibitem[1992]{gauvin92} 
Gauvin, L.~S.~\& Strom, K.~M.\ 1992, \apj \ 385, 217 

\bibitem[2000]{gizis00} 
Gizis, J.~E., Monet, D.~G., Reid, I.~N. et al. \ 2000, \apj \ 120, 1085 

\bibitem[2001]{gomez01} 
G\'omez, M.~\& Kenyon, S.~J.\ 2001, \aj \ 121, 974 

\bibitem[2002]{gomez02} 
G\'omez, M.~\& Persi, P.\ 2002, \aap \ 389, 494 

\bibitem[2003]{gomez03} 
G\'omez, M.~\& Mardones, D.\ 2003, \aj \ 125, 2134 

\bibitem[1989]{gh89} 
Gregorio-Hetem, J.~C., Sanzovo, G.~C., \& Lepine, J.~R.~D.\ 1989, 
\aaps \ 79, 452 

\bibitem[1992]{hartigan92} 
Hartigan, P.\ 1992, Bulletin of the American Astronomical Society 24, 1262 

\bibitem[2001]{joergens01} 
Joergens, V.~\& Guenther, E.\ 2001, \aap \ 379, L9 

\bibitem[2003]{joergens03} 
Joergens, V., Fern\'andez, M., Carpenter, J.~M. \& Neuh\"auser, R.\ 2003, 
astro-ph/0305397 

\bibitem[1985]{jones85} 
Jones, T.~J., Hyland, A.~R., Harvey, P.~M. et al.\ 1985, \aj \ 90, 1191 

\bibitem[2001]{kenyon01} 
Kenyon, S.~J.~\& G\'omez, M.\ 2001, \aj \ 121, 2673 

\bibitem[1995]{kenyon95} 
Kenyon, S.~J.~\& Hartmann, L.\ 1995, \apjs \ 101, 117 

\bibitem[1991]{kirkpatrick91} 
Kirkpatrick, J.~D., Henry, T.~J., \& McCarthy, D.~W.\ 1991, \apjs \ 77, 417 

\bibitem[1999]{kirkpatrick99} 
Kirkpatrick, J.~D., Reid, I.~N., Liebert, J.,~et al.\ 1999, \apj \ 519, 802 

\bibitem[2000]{kirkpatrick00} 
Kirkpatrick, J.~D., Reid, I.~N., Liebert, J.,~et al.\ 2000, \aj \ 120, 447 

\bibitem[2003]{lamm03} 
Lamm, M., Bailer-Jones, C.~A.~L., Mundt, R. et al., in preparation

\bibitem[1992]{landolt92} 
Landolt, A.~U.\ 1992, \aj \ 104, 340 

\bibitem[1996]{lawson96} 
Lawson, W.~A., Feigelson, E.~D., \& Huenemoerder, D.~P.\ 1996, 
\mnras \ 280, 1071 

\bibitem[2002]{lm02} 
L{\' o}pez Mart{\' i}, B., Stelzer, B. \& Neuh\"auser, R.\ 2002, 
\emph{AG Abstracts Series} 19, P80 

\bibitem[2000]{lucas00} 
Lucas, P.~W.~\& Roche, P.~F.\ 2000, \mnras \ 314, 858. 

\bibitem[2001]{lucas01} 
Lucas, P.~W., Roche, P.~F., Allard, F., \& Hauschildt, P.~H.\ 2001, 
\mnras \ 326, 695 

\bibitem[1997]{luhman97} 
Luhman, K.~L., Liebert, J. \& Rieke, G.~H. 1997, \apj \ 489, L165 

\bibitem[1999]{luhman99} 
Luhman, K.~L. 1999, \apj \ 525, 466 

\bibitem[1979]{luyten79} 
Luyten, W.~J.\ 1979, Minneapolis: University of Minnesota, 1979, 2nd ed. 

\bibitem[1996]{martin96} 
Mart{\' i}n, E.~L., Rebolo, R. \& Zapatero Osorio, M.~R.\ 1996, 
\apj \ 469, 706 

\bibitem[1999]{martin99} 
Mart{\' i}n, E.~L., Delfosse, X., Basri, G.,~et al.\ 1999, \aj \ 118, 2466 

\bibitem[2001]{martin01} 
Mart{\' i}n, E.~L., Dougados, C., Magnier, E., et al. 2001, \apjl \ 561, L195 

\bibitem[1999]{mizuno99} 
Mizuno, A., Hayakawa, T., Tachihara, K.,~et al.\ 1999, \pasj \ 51, 859 

\bibitem[2002]{mokler02}
Mokler, F. \& Stelzer, B. 2002, \aap \ 391, 1025

\bibitem[2003]{moraux03}
Moraux, E., Bouvier, J., Stauffer, J.~R. \& Cuillandre, J.-C. 2003, 
\aap \ 400, 891 

\bibitem[2001]{morrison01} 
Morrison, J.~E., R\"oser, S., McLean, B. et al. \ 2001, \aj \ 121, 1752 

\bibitem[2001]{natta01} 
Natta, A.~\& Testi, L.\ 2001, \aap \ 376, L22 

\bibitem[1998]{neuhauser98} 
Neuh\"auser, R.~\& Comer\'on, F.\ 1998, \emph{Science} 282, 83 

\bibitem[1999]{neuhauser99} 
Neuh{\"a}user, R.~\& Comer{\'o}n, F.\ 1999, \aap \ 350, 612 

\bibitem[2002]{neuhauser02} 
Neuh{\" a}user, R., Brandner, W., Alves, J., et al. 2002, \aap \ 384, 999 

\bibitem[1999]{oasa99} 
Oasa, Y., Tamura, M., \& Sugitani, K.\ 1999, \apj \ 526, 336 

\bibitem[2002]{oliveira02} 
Oliveira, J.~M., Jeffries, R.~D., Kenyon, M.~J. et al.\ 2002, \aap \ 382, L22 

\bibitem[2003]{oliveira03} 
Oliveira, J.~M., Jeffries \& van Loon, J.~Th. 2003, astro-ph/0310254

\bibitem[2000]{persi00} 
Persi, P., Marenzi, A.~R., Olofsson, G.,~et al.\ 2000, \aap \ 357, 219 
         
\bibitem[2000]{pickett00} 
Pickett, B.~K., Durisen, R.~H., Cassen, P., \& Mejia, A.~C.\ 2000, 
\apjl \ 540, L95
         
\bibitem[1991]{prusti91} 
Prusti, T., Clark, F.~O., Whittet, D.~C.~B. et al. 1991, \mnras \ 251, 303 
 
\bibitem[2000]{reid00} 
Reid, I.~N., Hawley, S.~L. 2000, 
\emph{New Light on Dark Stars}, Springer-Praxis
 
\bibitem[2001]{reipurth01} 
Reipurth, B.~\& Clarke, C.\ 2001, \aj \ 122, 432 

\bibitem[1993]{reipurth93} 
Reipurth, B.~\& Zinnecker, H.\ 1993, \aap \ 278, 81 

\bibitem[1985]{rieke85} 
Rieke, G.~K. \& Lebofsky, M.~J.\ 1985, \apj \ 288, 618 

\bibitem[1986]{robin86} 
Robin, A.~\& Crez\'e, M.\ 1986, \aap \ 157, 71 

\bibitem[1977]{schwartz77} 
Schwartz, R.~D. 1977, \apjs \ 35 \ 161
        
\bibitem[1991]{schwartz91} 
Schwartz, R.~D. 1991, in: Reipurth, B. (Ed.), 
ESO Scientific Report No.11,  
\emph{Low Mass Star Formation in Southern Molecular Clouds}, 93

\bibitem[1987]{stetson87} 
Stetson, P.~B.\ 1987, \pasp \ 99, 191 

\bibitem[1999]{sterzik} 
Sterzik, M.~F. \& Durisen, R.H. 1999, 
in: T. Nakamoto (Ed.), \emph{Proc. Star Formation}, p.387

\bibitem[2002]{tej02} 
Tej, A., Sahu, K.~C., Chandrasekhar, T., \& Ashok, N.~M. 2002, \apj 578, 523

\bibitem[1997]{whittet97} 
Whittet, D.~C.~B., Prusti, T., Franco, G.~A.~P. et al.\ 1997, \aap \ 327, 1194 

\bibitem[1999]{wilking99} 
Wilking, B.~A., Greene, T.~P., \& Meyer, M.~R.\ 1999, \aj \ 117, 469 
        
\bibitem[1997]{zo97} 
Zapatero Osorio, M.~R., Mart\'{\i}n, E.~L. \& Rebolo, R. 1997, \aap \ 323, 105 

\bibitem[2002a]{zo02a} 
Zapatero Osorio, M.~R., B{\' e}jar, Mart\'{\i}n, E.~L. et al.\ 2002, 
\apj 569, L99

\bibitem[2002b]{zo02} 
Zapatero Osorio, M.~R., B{\' e}jar, V.~J.~S., Pavlenko, Y., et al.\ 2002, 
\aap \ 384, 937 

\end{thebibliography}
\end{document}